\documentclass[
reprint,
groupedaddress,
 amsmath,amssymb,
 aps,
prb,
]{revtex4-1}

\usepackage{graphicx}
\usepackage{epstopdf}
\usepackage{dcolumn}
\usepackage{bm}

\begin{document}

\preprint{APS/123-QED}

\title{Atomistic calculation of the thermoelectric properties of Si nanowires}

\author{I.~Bejenari$^{1,2}$}
 \email{ igor.bejenari@fulbrightmail.org}
\author{P.~Kratzer$^1$}%
 \email{Peter.Kratzer@uni-due.de}
\affiliation{
${^1}$ Fakult\"at f\"ur Physik and Center for Nanointegration (CENIDE), Universit\"at Duisburg-Essen, 47048 Duisburg, Germany \\
${^2}$ Institute of Electronic Engineering and Nanotechnologies, Academy of Sciences of Moldova, MD 2028 Chisinau, Moldova}
\date{\today}
\begin{abstract}
The thermoelectric properties of 1.6 nm-thick Si square nanowires with $[100]$ crystalline orientation are 
calculated over a wide temperature range from 0~K to 1000~K,  taking into account atomistic electron-phonon interaction. In our model, the $[010]$ and $[001]$ facets are passivated by hydrogen and there are Si-Si dimers on the nanowire surface. The electronic structure was calculated by using the ${sp^3}$ spin-orbit-coupled atomistic second-nearest-neighbor tight-binding model. The phonon dispersion was calculated 
from a valence force field model of the Brenner type.

A scheme for calculating electron-phonon matrix elements from a second-nearest neighbor tight-binding model is presented. 
Based on Fermi's golden rule, the electron-phonon transition rate was obtained by combining the electron and phonon eigenstates. Both elastic and inelastic scattering processes are taken into consideration. The temperature dependence of transport characteristics was calculated by using a solution of linearized Boltzmann transport equation obtained by means of the iterative Orthomin method. At room temperature, the electron mobility is 195~cm$^{2}$V$^{-1}$s$^{-1}$ and increases with temperature, while a figure-of-mertit ZT=0.38 is reached for n-type doping with a concentration of $n=10^{19}$~cm$^{-3}$. 
\end{abstract}

\pacs{73.63.Nm, 73.50.Lw}
\maketitle
\section{\label{sec:level1} INTRODUCTION}

Silicon nanowires (NW) represent building blocks for nanoscale electronics.\cite{Cui_Science2001} They can be fabricated with a very good control of composition, size and shape.\cite{Ma_Science2003, Wu_NanoLett2004, Schmidt_AdvMater2009} For Si NW with a diameter of 5 nm and less, the number of atoms in the NW cross section becomes countable. Therefore, one should take into account electron-phonon interaction, crystalline orientation and quantum confinement to estimate transport characteristics of Si NW in order to predict the performance of nanoscale transistors, sensors and thermoelectric devices. In contrast to the bulk materials conventionally used in thermoelectrics, the nanostructured materials offer a possibility to design thermoelectric devices with an improved efficiency by exploiting the quantum confinement of electrons and phonons on the nanoscale.\cite{Boukai_Nature2008, Hochbaum_Nature2008} In this case, a fully atomistic simulation considering both the electron and phonon band structures as well as electron-phonon interaction is required to estimate thermoelectric properties.
 
The calculation of the transport properties of nanowires poses special challenges. Both the electronic and the vibrational spectrum are strongly altered compared to bulk by confinement effects.
The effective mass approximation and the k.p theory were widely used for transport calculations.\cite{Dresselhaus_PRB2000,Bejenari_PRB2008,Ramayya_JAP2008,Shin_JAP2009} These approaches become questionable for nanowires, at least if their diameter is just a few lattice constants. 
 Atomistic methods, such as the tight-binding (TB) electronic structure approach or density functional theory calculations, should be used instead.\cite{Neophytou_PRB245305,Buin_JAP2008,Rurali_RMP2010}
In Refs.~\onlinecite{Neophytou_PRB245305} and~\onlinecite{Buin_JAP2008}, the electron-phonon coupling was described by using the deformation potential approach. 
However, care must be taken when carrying over 
bulk-derived deformation potentials to nanowires.  
The classification of the phonons into acoustic and optical modes is helpful for bulk modes, but a more subtle classification (including flexural and torsional modes) is required for the vibrational modes of a nanowire.\cite{Mizuno_JPCM2009} For these classes, deformation potentials are not available.  Moreover, empirical data for deformation potentials are applicable only for the lowest (bulk) conduction band, and for a limited temperature range, usually around room temperature.

Theoretical work properly addressing the atomistic details of electron-phonon coupling in nanowires are still scarce. 
For example, Yamada {\it et al.}\cite{Yamada_JAP2012} calculated the electron mobility for different diameters and growth directions of Si NWs taking into account the electron-phonon scattering derived in the framework of the first nearest-neighbor ${sp^{3}d^{5}s^{*}}$ TB model. The Si dimers  on the $[010]$ and $[001]$ Si NW surfaces were not considered. The calculations were done considering only the lowest conduction subbands at a room temperature. At high temperatures, this treatment is not adequate because the high energy electron subbands play essential role in the transport properties. 
Using the similar treatment of the the electron-phonon scattering,  Zhang {\it et al.}\cite{Zhang_PRB2010} calculated the electron mobility for $[110]$-oriented Si NWs with different diameters. Their calculations were done at two temperatures, 77~K and 300~K. These data are insufficient to draw firm  conclusions about the temperature dependence of the electron mobility.

Here, our aim is to investigate the use of Si NWs for thermoelectrics over a wide temperature range. This requires to study the temperature dependence of electron-phonon scattering in the NWs. In addition to the temperature dependence of the scattering rate, it is also important to account for its dependence on electron energy. This is because, in thermoelectrics, electrons with energies considerably above the band bottom may have a significant impact on the Seebeck coefficient and the thermal conductivity. This is on contrast to the ohmic regime of charge transport, where electrons just above  the band edge dominate the transport properties.

The focus of this work is to calculate accurately transport properties of Si NWs based on an atomistic model, which is close to a real system. For this purpose, we take into account relaxation in the atomistic structure, presence of Si dimers on the $[010]$ and $[001]$ Si NW surfaces, all confined phonon modes, and all electron subbands in an energy range from the conduction band bottom to 5${k_B T}$ above. To calculate the matrix elements of the electron-phonon coupling Hamiltonian, we have used the tight-binding approach for the electronic structure, employing the phononic structure obtained from an atomistic force field method as input.

The rest of the paper is organized as follows. In Sec. II, we discuss the electronic and phonon band structures as well as the  underlying physics. In Sec. III, we describe thermoelectric transport coefficients in the framework of the linearized Boltzmann transport theory. Also, we describe the electron-phonon transition rate in terms of the TB formalism. We discuss the energy and temperature dependence of the transport distribution function and total scattering rate as well as the relaxation time as a solution of the Boltzmann transport equation. The temperature dependencies of the thermoelectric parameters and electron mobility of \textit{n}-type Si NWs are presented in Sec. IV. Finally, our conclusions are given in Sec. V. In Appendices A -- C, a detailed description of the matrix elements of the electron-phonon coupling Hamiltonian in terms of the $sp^3$ 2nd nearest-neighbor TB formalism is presented. We also describe the iterative Orthomin(1) method used for solving the Boltzmann equation.  

\section{Band Structure Calculation}

\subsection{Model}

\begin{figure}[h]
\includegraphics[width=7cm]{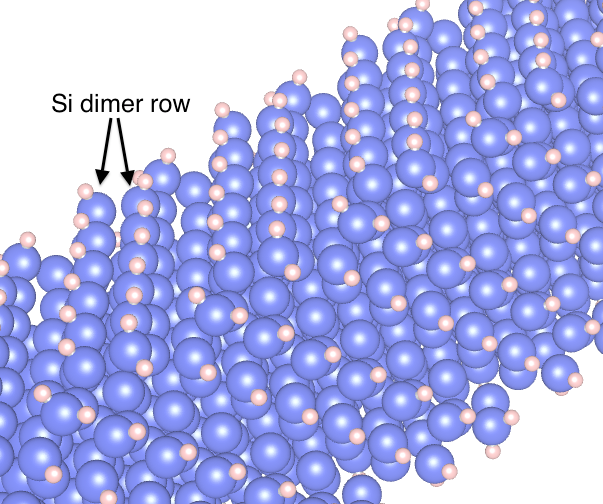}
\caption{\label{fig:fig1} (Color online) Schematic diagram representing the Si NW structure passivated by hydrogen atoms. The Si (H) atoms are represented by large (small) spheres.}
\end{figure}

Figure~\ref{fig:fig1} shows the  Si NW structure with square-shaped cross-section along the $[100]$ crystalline orientation. The NW length is assumed to be infinite in the transport direction ${x}$. The NW thickness is 1.6 nm. 
The Si-Si dimers are shown on the lateral NW surface. To properly take into account these dimers, we considered the supercell with a length of ${2a}$, where ${a=5.429 ~\text{\AA}}$ is a lattice parameter of the bulk silicon. The width of the NW is ${3a}$. The surface Si atoms with one dangling bond are hydrogenated. In the non-relaxed structure, the Si-Si and Si-H bond lengths are 2.35~\AA~and 1.48~\AA, respectively.\cite{Guzman-Verri_JPCM2011,Bacalis_JMC2009} Using  the General Utility Lattice Program (GULP) based on force field methods, a relaxation of atom positions about the given atom coordinates was achieved by means of a minimization of the total energy of the atomic system.\cite{Gale_MS2003,Gale_JPCB1998} Both Rational Functional Optimization and Conjugate Gradients methods were used to calculate relaxation in the atomistic structure.

We have considered ${n}$-type Si nanowires doped by phosphorus atoms. The donor charge transfer level is equal to ${E_{D}=-0.045}$~eV relative to the conduction band edge.\cite{Madelung_2004,Ashcroft_1976} Here, we do not take into consideration the electron-impurity scattering. 
The impurity concentrations quoted in the results are merely used to define the temperature dependence of the Fermi level.

\subsection{Electronic Structure}

The electronic band structure for Si NWs was computed using a semi-empirical TB approach, where the two-center orthogonal ${sp^3}$ model was used taking into account the 1st and 2nd nearest neighbors.\cite{Grosso_PRB1995,Santoprete_2004} 
The tight-binding parameters were chosen in such a way to accurately represent the band gap of bulk silicon. 
The model allows for a dependence of the tight-binding matrix elements on both the bond angles and on the bond distances, using power laws, with exponents chosen to reproduce the known deformation potentials of bulk silicon. 
The matrix elements of the TB Hamiltonian are presented in Appendix A. 
In order to enable an accurate description of the effects of crystal deformations on the electronic structure, we include effects of 'screening" of the interaction with the 2nd-nearest neighbor atom due to displacements of the nearest neighbor atom.
To calculate the eigenvectors and eigenvalues of the Hamiltonian, we have used the standard library LAPACK.\cite{Barker_2001}

\begin{figure}[h]
\includegraphics[width=7cm]{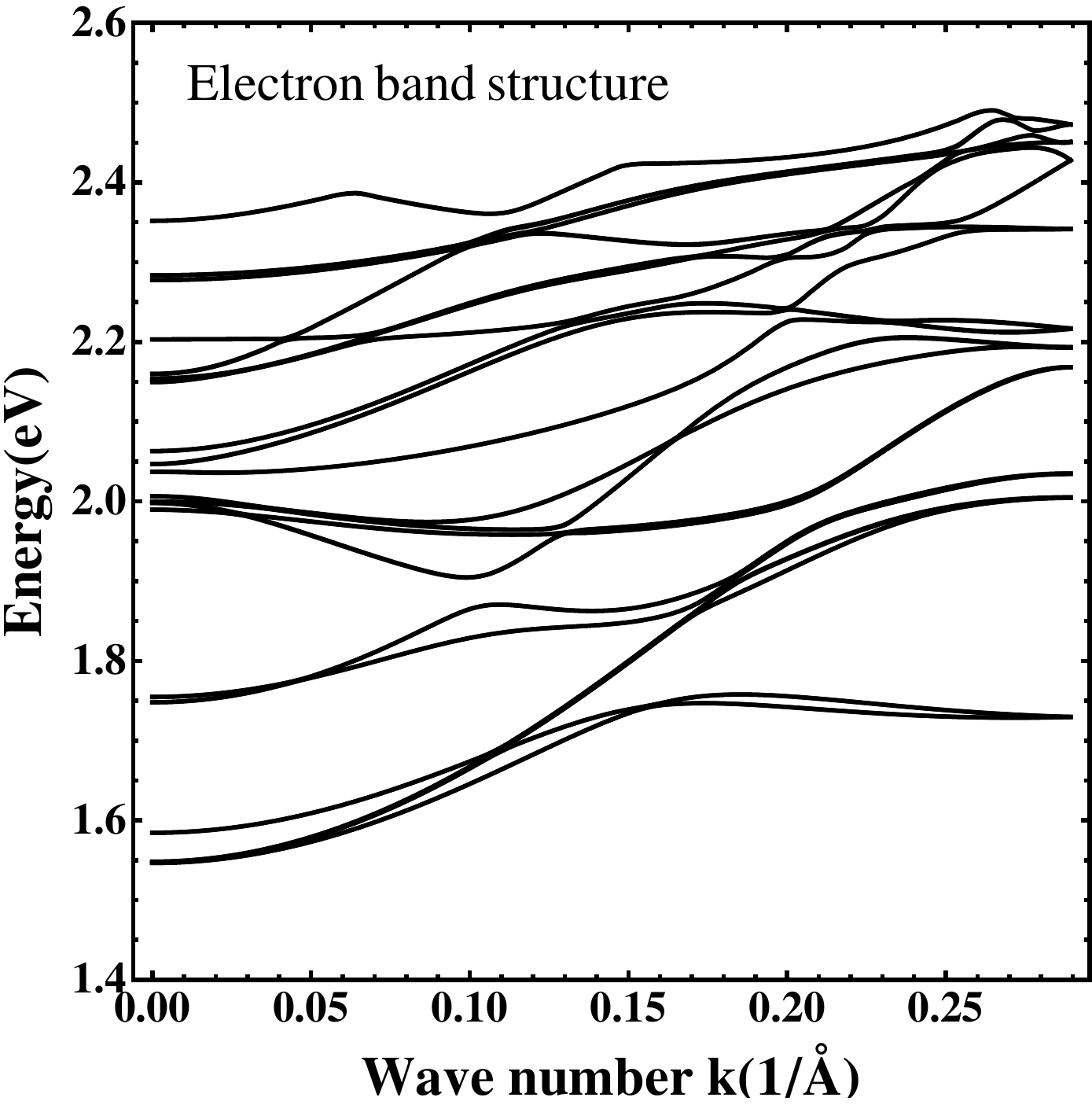}
\caption{\label{fig:fig2} The electron band structure computed for the [100]-oriented Si NW with a thickness of 1.6 nm. The graph includes 40 subbands, including degenerate ones. The zero of energy is set to the valence band top of bulk Si.}
\end{figure}

In the Si bulk material, the conduction band structure consists of six valleys placed along the lines of ${\Delta_{1}}$ symmetry close to the ${X}$ point in the Brillouin zone. It is an indirect band gap semiconductor, because the edges of the valence bands are located at the ${\Gamma}$ point. Its band gap is equal to 1.12 eV~(Ref.~\onlinecite{Grosso_PRB1995}). Figure~\ref{fig:fig2} represents the electron band structure computed for the [100]-oriented Si NW with a thickness of 1.6 nm. The origin of the energy axis corresponds to the top of the valence band at the ${\Gamma}$ point in the Brillouin zone of the bulk Si material.  
The spin-orbit (SO) interaction is rather weak in the Si material, hence, the energy subband splitting due to the SO effect is invisible in the plot, and each plotted line corresponds to a doubly degenerate state. 
For the given NW growth orientation, the six electron ellipsoidal valleys equivalent in bulk Si are split into two groups, including four and two ellipsoids in the Si NW, respectively. 
The confinement effect and the symmetry lowering due to surface dimerization lead to a further splitting of the conduction subbands. At the ${\Gamma}$ point, the bottoms of the conduction subbands are 1.546 (doubly degenerate), 1.548 and 1.584 eV, respectively.
These subbands originate from the backfolding of the four ellipsoids located in the $k_y k_z$-plane onto the $\Gamma$ point because of the confinement in the spatial directions perpendicular to the NW axis. They show a rather large dispersion at $\Gamma$ as functions of $k_x$, which corresponds to a direction of low effective mass of these ellipsoids.  
Two of these subbands reach the edge of the Brillouin zone at $k_x = 0.289 \, \text{\AA}^{-1}$ at an energy of 1.72~eV. 
One of them even has a shallow minimum at  0.271~${\text{\AA}^{-1}}$. 
These two subbands change their orbital character as function of $k_x$, and near the Brilloiun zone boundary they obtain the character of the states in the two ellipsoids along the (100)-direction in bulk Si. As these ellipsoids have their heavy-mass direction along (100), the subbands show a weak disperison and are rather flat at the Brillouin zone boundary. 
The Si NW is a direct bandgap semiconductor. Its band gap is found to be equal to 2.18 eV at the ${\Gamma}$ point. 
Because of the different effective masses, one can expect that the transport properties of ${n}$-type Si NWs are mainly defined by the electrons in the band minima centered at the ${\Gamma}$ point rather than by the 
electronic states near the Brillouin zone boundary. 

A comparison of our results with data obtained by Markussen {\it et al.} indicates that the presence of dimers on the NW surface leads to a significant modification of the electronic structure. For example, the electron density of states (DOS) increases at the ${\Gamma}$ point.\cite{Markussen_PRB2009}  In the transport calculations, we considered 12 electron subbands fitted in the energy range of 5${k_B T}$. At the ${\Gamma}$ point, spanning the energy interval from 1.5 to 1.8 eV.

\subsection{Phonon Band Structure}

\begin{figure}[h]
\includegraphics[width=7cm]{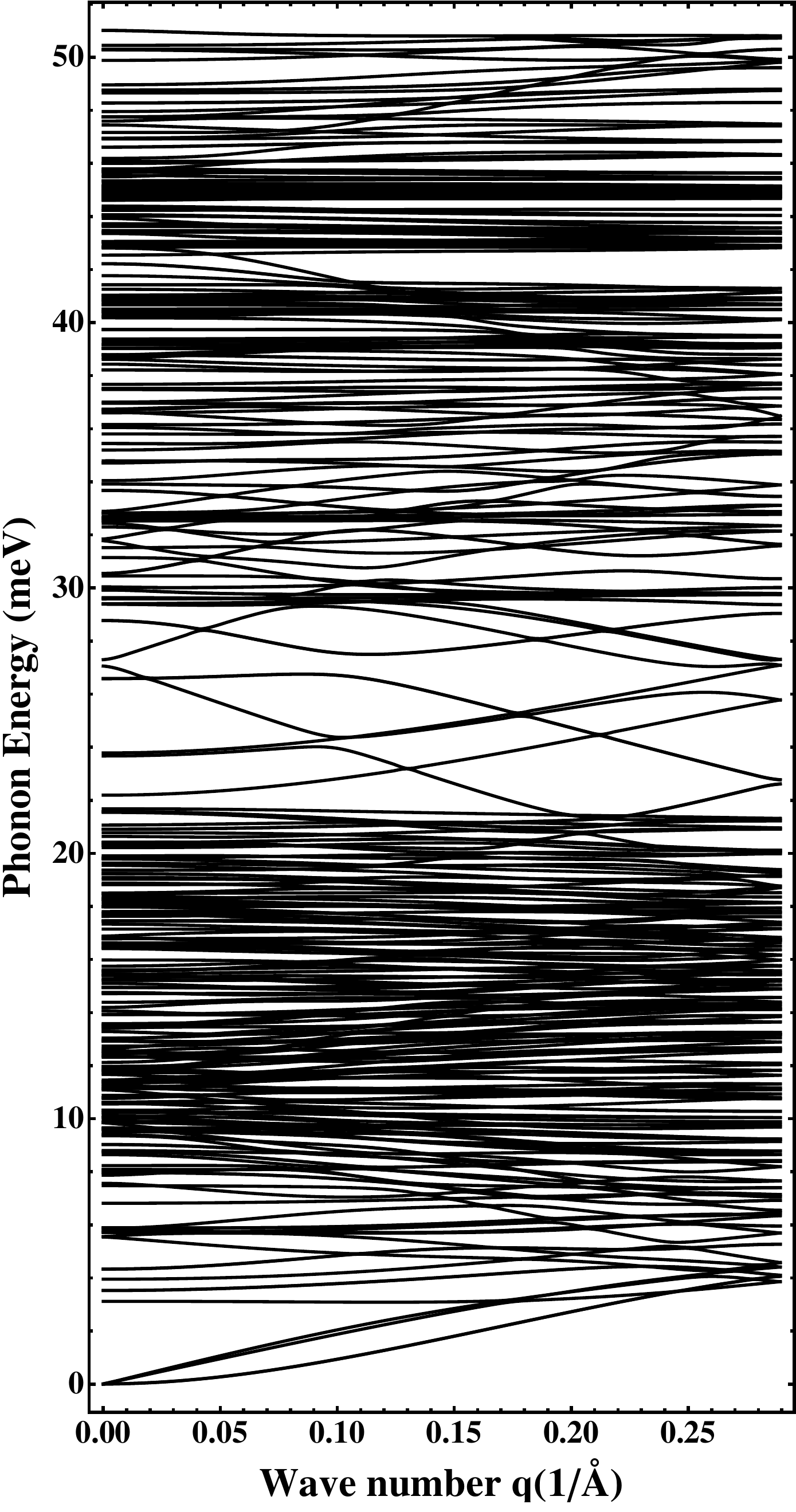}
\caption{\label{fig:fig3} The phonon band structure computed for the [100]-oriented Si NW with a thickness of 1.6 nm. The graph includes 410 phonon modes. The energy of the Si--H stretch modes is outside the scale of the graph.}
\end{figure}

Based on an atomistic model, the phonon band structure  was calculated in the framework of Brenner's valence force field model. The eigenvalues and eigenvectors of the dynamical matrix were computed by using GULP.\cite{Gale_MS2003,Gale_JPCB1998}  Figure~\ref{fig:fig3} shows the phonon band structure for the [100]-oriented Si NW with a thickness of 1.6 nm. In this system of 226 atoms, there are 678 different phonon modes. The phonon modes with the highest excitation energy correspond to the very light hydrogen atoms. Their energies belong to the interval from 270 to 285 meV. These modes have no effect on the transport properties. The phonon density of states is a non-monotonic function of energy. It has a clear minimum in the energy range between 22 and 29~meV. The lower boundary of this interval reflects the highest possible energy of the transverse acoustic mode in bulk Si. 
The phonon DOS is sharply-peaked around energy values of 14, 39, and 58 meV. This results from mixed states of acoustic and optical phonons, as well as the optical phonon modes. The non-monotonic dependence of the phonon DOS effects the temperature dependence of the transport coefficients described in the following sections.

In contrast to the bulk phonon dispersion, there are four soft modes with $\omega(q) \approx 0$ in a long wave-length regime in NWs. These are dilatational (longitudinal), flexural, shear and torsional (transverse) phonon acoustic modes. The doubly degenerate flexural mode represents a mixed state of transverse and longitudinal phonon acoustic modes. At low temperature, the acoustic modes are mainly responsible for the electron-phonon scattering. Based on the results of the numerical calculations, we have introduced functions fitting the dispersion law of the acoustic modes. A comparison between our results and the dispersion law of the soft acoustic modes obtained by Mizuno {\it et al.}\cite{Mizuno_JPCM2009} by means of the continuous medium approximation based on group theory is presented in Table~\ref{tab:table1}. 
The continuous medium approximation would predict a quadratic behavior on the wave vector $q$ while our calculations yields a smaller exponent. 
The difference in the dispersion laws is mainly due to the fact that the continuous medium approximation does not take into consideration the surface elastic energy of NWs. 
We convinced ourselves that the exponent of the dispersion law for the flexural modes increases with increasing NW thickness. For example, it is equal to 1.88 for a NW thickness of 2.7 nm. For the dilatational phonon modes, the dispersion law obtained in different models is the same, because the lateral NW sides do not influence the longitudinal elastic waves propagation.
\begin{table}[h]
\caption{\label{tab:table1}
Dispersion law of the soft acoustic phonon modes in the long wavelength  regime in a square Si NW.} 
\begin{ruledtabular}
\begin{tabular}{lcccccccc}
Acoustic mode & degeneracy & ${\omega(q)}$\footnote{Present work.} & ${\omega(q)}$\footnote{Reference ~[\onlinecite{Mizuno_JPCM2009}].} \\
\hline
Dilatational & 1 & ${\approx q}$ & ${\approx q}$  \\
Flexural & 2 & ${ q^{1.28}}$   & ${ q^{2}}$ \\
Torsional & 1 & ${\sqrt{\omega^{2}_0 +v^{2}_{t} q^{2}}} $ \footnote{${\omega_0=1.2\times10^{-5}}$ eV and ${v_{t}=523}$ m/s for a NW thickness of 1.6 nm.}  & ${q}$ \\
\end{tabular}
\end{ruledtabular}
\end{table}
 A comparison of the calculated displacement fields of the phonon modes with the results reported by Mizuno {\it et al.} indicates that at least the lowest ten phonon modes are considered to be purely acoustic ones.
There are a single dilatational, six flexural, two shear and one torsional acoustic phonon modes. The soft modes reported in Table~\ref{tab:table1} correspond to the acoustic phonon modes with the lowest energy, i.e., in a long wave-length regime.
 The higher-lying phonon modes represent mixed states of acoustic and optical phonons. We note that the dispersion law for the torsional phonon mode found in our calculation agrees with that obtained by Buin {\it et al.} from the elastic wave equation and given by the Pochhammer-Chree equation for cylindrical Si NWs.\cite{Buin_JAP2008}

\section{Atomistic Boltzmann theory}

\subsection{Thermoelectric related transport coefficients}

In the linearized Boltzmann formalism (see e.g. Ref.~\onlinecite{Ashcroft_1976}), the transport coefficients, i.e., electrical conductivity (${\sigma}$), Seebeck coefficient (\textit{S}) and thermal conductivity (${\kappa}$) are defined in terms of the moments $L^{(\alpha)}$ of the distribution function as
\begin{eqnarray}
\sigma &=& e^{2}L^{(0)}, \\
S &=& - \frac{1}{eT} \frac{L^{(1)}}{L^{(0)}}, \\
\label{eq:Seebeck}
\kappa_{el} &=& \frac{1}{T} \left[ L^{(2)} -\frac{ \left(L^{(1)}\right)^{2}}{L^{(0)}} \right].
\label{eq:kappa_el}
\end{eqnarray}
The dimensionless figure of merit ${ZT}$ that characterizes the efficiency of a thermoelectric material is defined as\cite{Nolas_2001}  
\begin{equation}
 ZT = \frac{\sigma S^{2}}{\kappa} T.
\end{equation}
When calculating $ZT$, we use the full thermal conductivity $ \kappa=\kappa_{ph} + \kappa_{el}$. 
For thin Si NWs, the phononic thermal conductivity ${\kappa_{ph}=7}$~WK$^{-1}$m$^{-1}$ (see Ref.~\onlinecite{Cruz_JAP2011}). The $\alpha$th moment of the distribution function reads 
\begin{eqnarray}
L^{(\alpha)} &=&  \frac{2}{\pi k_{B} T A} \sum_{n} \int_{0}^{\pi/2a}  v^{2}_{n}(k_{x}) \tau_{n}(k_{x}) \left[ E_{n}(k_{x}) -E_{F} \right]^{\alpha} 
 \nonumber \\
&&  \times 
f_{0}\left[E_{n}(k_{x}) \right] \left\lceil 1- f_{0}\left[E_{n}(k_{x}) \right] \right\rceil  dk_{x} 
\end{eqnarray}
where ${A}$ is the cross-sectional area of the NW, ${\tau_{n}(k_{x})}$ is the relaxation time,  
${f_{0}(E)}=1/ \left\lbrace\exp[(E-E_{F}) / k_{B} T] +1 \right\rbrace $ is the equilibrium Fermi-Dirac distribution function and ${v_{n}(k_{x})= \partial E_{n}(k_{x}) / \hbar \partial k_{x}}$ is the group velocity of the electrons. The upper limit of integration is ${\pi/(2a)}$ because the length of the supercell  is ${2a}$ along the ${x}$-axis. The moments of  distribution function can be written   in terms of the transport distribution function ${\Xi(E)}$ as\cite{Mahan_PNAS1996,Scheidemantel_PRB2003}
\begin{equation}
L^{(\alpha)} =  \int \left(- \frac{\partial f_{0}}{\partial E} \right) \Xi(E) \,  \big( E -E_{F} \bigr)^{\alpha}  dE 
\label{eq:Lalpha_TDF}
\end{equation}
The transport distribution function (TDF) is defined as\cite{Neophytou_PRB245305} 
\begin{equation}
 \Xi(E) = \frac{1}{A} \sum_{n, k_{x}} v^{2}_{n}(k_{x}) \tau_{n}(k_{x}) \delta \bigl( E-E_{n}(k_{x}) \bigr).
\end{equation}
Being the integral kernel of all transport coefficients, the TDF provides the full information about the transport processes. It plays a similar central role as the transmission probability in the Landauer formalism of coherent transport in the linear response regime.\cite{Jeong_JAP2010}
The TDF depends on the electron energy both through the scattering rate and the group velocity. 

The electron mobility ${\mu}$ is evaluated from the relationship ${\sigma=en\mu}$. The electron density $n$ is computed from the following expression
\begin{equation}
 n= \frac{2}{\pi A} \sum_{n}  \int_{0}^{\pi/2a} f_{0}\left( E_{n}(k_{x}) \right) dk_{x}.
 \end{equation}
To calculate the temperature dependence of the transport coefficients, one has to obtain the dependence of the Fermi energy on temperature when evaluating the moments of the distribution function. Similar to Ref.~\onlinecite{Fomin_PRB2010}, we obtain this dependence by solving the electroneutrality equation defined as  
\begin{equation}
 n(E_{F}) = n_{D} \frac{1}{2 \exp[(E_{F} -E_{c} -E_{D}) / k_{B} T] +1},
\end{equation}
where ${E_{c}}$ is the bottom of the  conduction band and ${n_{D}}$ is the concentration of donor atoms. Our calculations show that the temperature dependence of the Fermi energy in a Si nanowire is similar to that in the n-type Si bulk material.\cite{Wolfe_1989} $E_{F}$ moves towards ${E_{D}}$ as the temperature is lowered and the electrons in the conduction subbands freeze out. When the temperature increases, the Fermi energy tends towards the center of the band gap.

\subsection{Electron-phonon scattering}

Generally speaking, the electron-phonon coupling matrix elements may be obtained by a Taylor expansion in terms of the atomic displacements of the potential felt by an electron in a specific Bloch state (see Ref.~\onlinecite{Resta_PRB1991} and references therein). 
For the tight-binding formalism, this prescription translates into a calculation of the spatial derivatives of the Hamiltonian matrix elements. 
Subsequently, the scalar product of this gradient vector with each phonon eigenvector ${\vec{\epsilon}_{i,\lambda}}$ (polarization vecor) must be  evaluated, and finally all phonon modes are summed over to obtain the matrix element ${\cal M}$. 
This is in contrast to the conventional theory of deformation potentials in bulk materials which, as an  additional approximation, involves an expansion of the displacement fields in terms of a wave vector $q$, keeping only the leading non-vanishing order of the electron-phonon interaction in $q$ (the first order in $q$ for acoustic phonons, the zeroth order for optical phonons). In an attempt to take over this approach to structures with reduced spatial dimensions, its inherent deficiencies may be partially cured by applying selection rules; for an example pertinent to Si NWs see Ref.~\onlinecite{Neophytou_PRB085313}. 
These rules are used to decide which electronic states in a nanostructure are supposed to be coupled by the deformation potential without making use of information about the microscopic symmetries and orbital structure of the electronic wavefunctions involved. 
Contrarily, the approach presented in this work offers the advantage of an unbiased evaluation of the coupling matrix elements at the price of higher computational cost. 
The restrictions applied by the selection rules are implicitly included in our approach, as they automatically arise from the microscopic symmetries of the electron and phonon eigenvectors, as well as of the electron-phonon scattering potential.
 
We extend the formulas for the first nearest-neighbor ${sp^{3}d^{5}s^{*}}$ TB model described in Refs.~\onlinecite{Yamada_JAP2012} and \onlinecite{Zhang_PRB2010} to the more compact ${sp^3}$  second-nearest-neighbor TB model. In this TB model, a first-order expansion of the TB Hamiltonian as a function of the atomic displacements give us three contributions: the derivatives of 1st-nearest neighbor matrix elements ${\cal S}^{(1)}(i)$,  the derivatives of 2nd-nearest neighbor matrix elements w.r.t. 1st neighbor distances ${\cal S}^{(2)}_{1}(i)$, and w.r.t. 2nd neighbor distances ${\cal S}^{(2)}_{2}(i)$.  For the atomic site indices, we use the following convention: The four nearest neighbors of each site $i$ are enumerated by small roman indices, e.g. $j$, whereas capital roman indices, e.g. $M$ are used for the 12 next-nearest neighbors.  We obtain the following formula for the electron-phonon transition matrix element from an electron state ${(n,k_{x})}$  to a state ${(n',k'_{x})}$:  
\begin{widetext}
\begin{multline}
{\cal M}^{n,n'}_{\lambda}(k,k')= \sum_{K}^{0, \pm \pi/a} \sum_{q} \left( \frac{\hbar}{N_{at} \omega_{\lambda}(q)} \right)^{1/2} \delta_{k-k'\pm q, K}  \sum_{i}^{N_{at}} \sum_{O,O'}^{s,p_{x},p_{y}, p_{z}} \left( {\cal S}^{(1)}(i) + {\cal S}^{(2)}_1(i) + {\cal S}^{(2)}_2(i) \right)  \\ \times \exp \left\lbrace -i \left[ \frac{E_{n}(k)}{\hbar} - \frac{E_{n'}(k')}{\hbar} \pm \omega_{\lambda}(q) \right]t \right\rbrace ,
\end{multline}
where
\begin{eqnarray}
{\cal S}^{(1)}(i) &=& \sum_{j} \frac{\partial{ \langle O', \vec{R_{i}} | H | O, \vec{R}_{j} \rangle}^{(1)}}{\partial \vec{R}_{ij}} 
\nonumber\\  
& & \times
\left[ \frac{\vec{\epsilon}_{i,\lambda}(\pm q_{x})}{\sqrt{M_{i}}} \exp \left(i k'_{x} R_{x,ij} \right) - \frac{\vec{\epsilon}_{j,\lambda}(\pm q_{x})}{\sqrt{M_{j}}} \exp \left(i k_{x} R_{x,ij} \right) \right]   C_{i,O'}^{n' *}(k'_{x}) C_{j,O}^{n}(k_{x}), \\
\lefteqn{
{\cal S}^{(2)}_1(i)= \sum_{M} \frac{\partial{ \langle O', \vec{R_{i}} | H | O, \vec{R}_{M} \rangle}^{(2)}}{\partial \vec{R}_{jM}} } \nonumber\\  
& & \times \left[ \frac{\vec{\epsilon}_{j,\lambda}(\pm q_{x})}{\sqrt{M_{j}}} \exp \left(i k'_{x} R_{x,iM} \right) - \frac{\vec{\epsilon}_{M,\lambda}(\pm q_{x})}{\sqrt{M_{M}}} \exp \left(i k_{x} R_{x,iM} \right) \right]   C_{i,O'}^{n' *}(k'_{x}) C_{M,O}^{n}(k_{x}), \\
\lefteqn{
{\cal S}^{(2)}_2(i)= \sum_{M} \frac{\partial{ \langle O', \vec{R_{i}} | H | O, \vec{R}_{M} \rangle}^{(2)}}{\partial \vec{R}_{iM}} }\nonumber\\  
& & \times \left[ \frac{\vec{\epsilon}_{i,\lambda}(\pm q_{x})}{\sqrt{M_{i}}} \exp \left(i k'_{x} R_{x,iM} \right) - \frac{\vec{\epsilon}_{M,\lambda}(\pm q_{x})}{\sqrt{M_{M}}} \exp \left(i k_{x} R_{x,iM} \right) \right]   C_{i,O'}^{n' *}(k'_{x}) C_{M,O}^{n}(k_{x}),
\end{eqnarray}
\end{widetext}
where 
${C_{i,O}^{n}(k_{x})}$ is the expansion coefficient of the electronic wave function $(n,k_x)$ in terms of orbital $O$ at atiomic site $i$, 
${M_{i}}$ is the mass of the $i$th atom, 
$N_{at}$ is the number of atoms,  
and ${K}$ is a vector of the  reciprocal lattice, including the origin $K=0$. 
Both normal (${K=0}$) and umklapp (${K \neq 0}$) phonon scattering processes are included in the transition matrix element. The crystal momentum conservation law is taken into account by the Dirac delta function. 

The above formulas can be applied to both emission and absorption of a phonon with the frequency ${\omega_{\lambda}(q_{x})}$. The phonon and electron eigenvectors   satisfy to following relations ${\vec{\epsilon}_{i,\lambda}(- q_{x})=\vec{\epsilon}_{i,\lambda}^{*}(q_{x})}$ and ${C_{i,O}^{n}(-k_{x})=-i \sigma_{y} C_{i,O}^{n *}(k_{x})}$, correspondingly, where ${\sigma_{y}}$ is the Pauli matrix. In a case of spin degeneracy, the relation for the electron wave functions is simplified to ${C_{i,O}^{n}(-k_{x})= C_{i,O}^{n *}(k_{x})}$.The electron eigenenergies are implicitly considered by means of the  spatial derivatives of the Hamiltonian matrix elements presented in Appendix B. 

The transition rate of an electron from the initial state ${(n,k_{x})}$ to final state ${(n',k'_{x})}$ is given by applying Fermi's golden rule as follows:
\begin{widetext}
\begin{multline}
S^{n,n'}(k,k')=\frac{2 \pi}{\hbar} \sum_{\lambda} \left| {\cal M}_{\lambda}^{n,n'}(k,k') \right|^{2} g \left[ \hbar \omega_{\lambda}(k'-k) \right] \delta \left[ E_{n}(k) - E_{n'}(k') + \hbar \omega_{\lambda}(k'-k) \right] \\
+ \frac{2 \pi}{\hbar} \sum_{\lambda} \left| {\cal M}_{\lambda}^{n,n'}(k,k') \right|^{2} \left[1 +g \left( \hbar \omega_{\lambda}(k-k') \right) \right] \delta \left[ E_{n}(k) - E_{n'}(k') - \hbar \omega_{\lambda}(k-k') \right], \label{eq:Snn}
\end{multline}
\end{widetext} 
where ${g(\hbar \omega)}=1/ \left[\exp(\hbar \omega / k_{B} T) -1 \right]$ is the equilibrium Bose-Einstein phonon distribution function at the temperature $T$. The energy conservation law is included by the Dirac delta function. The first term in Eq.~(\ref{eq:Snn}) corresponds to the transition of an electron caused by the absorption of a phonon with the energy ${\hbar \omega_{\lambda}(q_{x})}$ and momentum  ${q_{x}=k'_{x}-k_{x}}$. The second term corresponds to the emission of a phonon with the energy ${\hbar \omega_{\lambda}(k_{x}-k'_{x})}$. Both quasi-elastic and inelastic electron-phonon scattering are taken into account in the transition rate. 

For the numerical evaluation of the formulas, we used cubic spline interpolation of the electron and phonon band structures between discretized points. We managed to reduce the CPU time for calculation of the transition rate by parallelizing our codes on multi-core computer architectures with OpenMP.\cite{openmp08} 

\subsection{Relaxation time}

The momentum relaxation time ${\tau_{n}(k_{x})}$ is calculated by numerically solving the integral Boltzmann equation of the form\cite{Yamada_JAP2012}
\begin{multline}
 \frac{L_{x}}{2\pi} \sum_{n'} \int_{-\pi/2a}^{\pi/2a} S^{n,n'}(k,k') \frac{1-f_{0}(E_{n'}(k'))}{1-f_{0}(E_{n}(k))}  
 \\
\times \left[ v_{n}(k) \tau_{n}(k) - v_{n'}(k') \tau_{n'}(\left|k'\right|) \right] dk' = v_{n}(k).
\label{eq:BEq}
\end{multline}
In the above equation, all in-scattering and out-scattering contributions of both the elastic and inelastic electron-phonon scattering are included in the integrand. The one-dimensional integration in  Eq.~(\ref{eq:BEq}) is reduced to the summation over a set of values ${ \left\lbrace k'_{r} \right\rbrace }$ by using the following property of the Dirac delta function 
\begin{multline}
\delta \left[ E_{n}(k) - E_{n'}(k') \pm \hbar \omega_{\lambda}(\pm k' \mp k) \right] = 
\\
\sum_{r} \frac{1}{\left| \left( \frac{\partial \left[ E_{n'}(k') \mp \hbar \omega_{\lambda}(\pm k' \mp k) \right] }{\partial k'} \right)_{k'=k'_{r}} \right|}  \delta(k'-k'_{r}), 
\label{eq:Dirac}
\end{multline}
where ${k'_{r}}$ are roots of the equation ${E_{n}(k) - E_{n'}(k') \pm \hbar \omega_{\lambda}(\pm k' \mp k) =0}$, which is the energy conservation law, at fixed values of ${n, n',\lambda}$ and ${k}$.

An iterative  solution of the reduced Boltzmann equation~(\ref{eq:BEq}) is obtained by means of the iterative Orthomin(1) method.\cite{Greenbaum_1997} This method is shortly described in Appendix C. As a seed of the iteration, we use the low-temperature relaxation time approximation\cite{Kawamura_PRB1992,Restrepo_APL2009}

\begin{equation}
 \frac{1}{\tau_{n}^{(0)}(k)} = \frac{L_{x}}{2\pi} \sum_{n'} \int_{-\pi/2a}^{\pi/2a} S^{n,n'}(k,k') \frac{1-f_{0}(E_{n'}(k'))}{1-f_{0}(E_{n}(k))}  dk'.
\label{eq:tauNull}
\end{equation}

\begin{figure}[h]
\includegraphics[width=7cm]{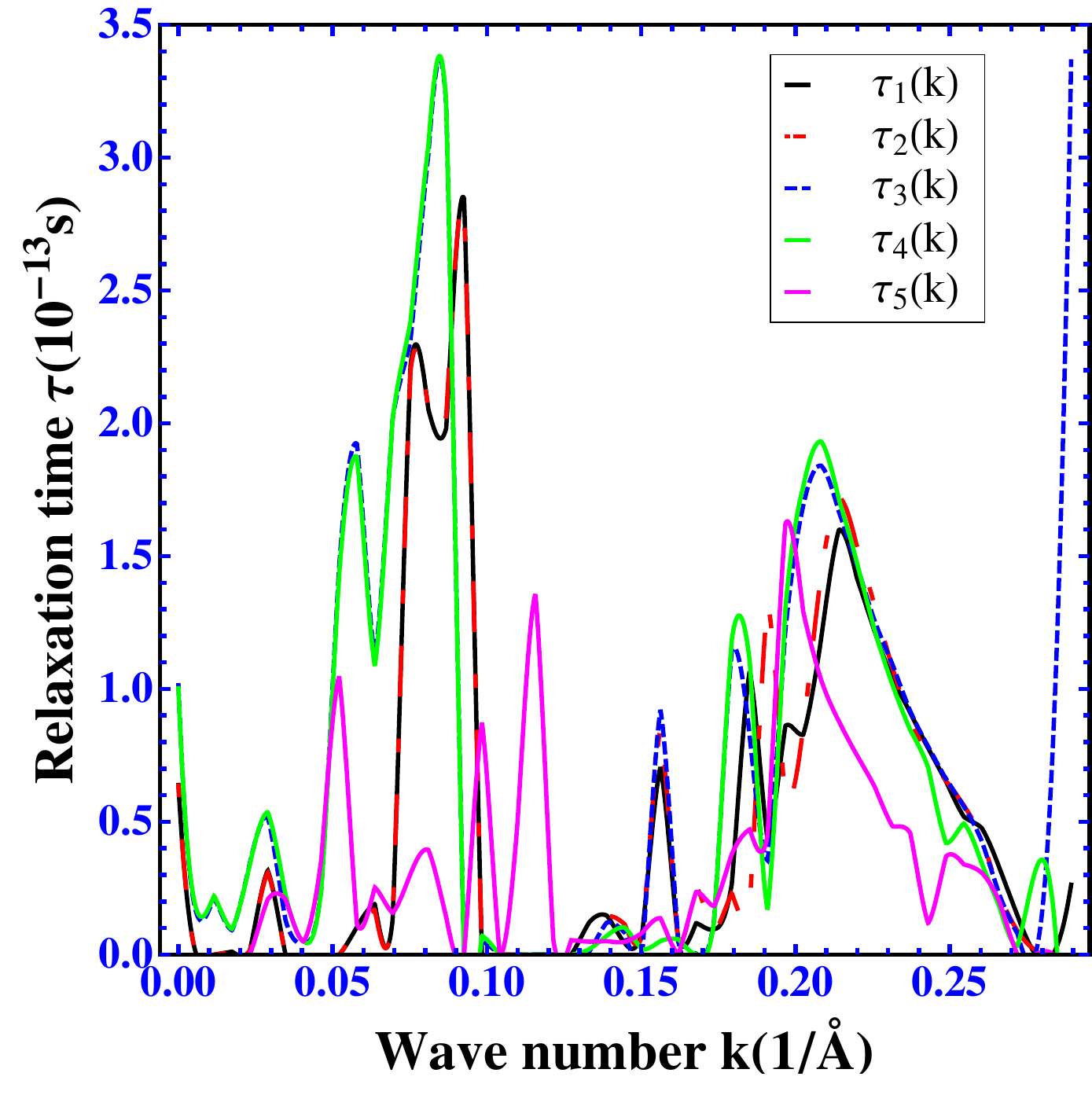}
\caption{\label{fig:fig4} (Color online) The momentum relaxation time ${\tau_{n}(k)}$ for the first five electron subbands computed for the [100]-oriented Si NW with a thickness of 1.6 nm at a room temperature.}
\end{figure}

Figure~\ref{fig:fig4} demonstrates the momentum relaxation time ${\tau_{n}(k)}$ for the first five electron subbands resulting from the numerical solution of Eq.~(\ref{eq:BEq}) for the Si NW. It reveals a strong variation of the relaxation time in momentum space, a behavior typical of transport in one-dimensional channels.\cite{Fomin_PRB2010} There are many kinks of ${\tau_{n}(k)}$ at the $k$-points corresponding to peculiarities in the electron band structure. The relaxation time increases at those $k$-points where the band dispersion is large, i.e., the electron group velocity is large. In this case, the electron-phonon interaction with acoustic phonons is small, because crystal momentum conservation severely restricts emission or absorption of such a phonon within the same electronic band. In contrast, the relaxation time is diminished in those intervals where the energetic distance between adjacent  electron subbands is small, or the electron subbands are flat or intersect with others. This leads to an increase in the electron-phonon scattering due to inter-band transitions. The 1st and 2nd (3rd and 4th) electron subbands are almost degenerate due to the very small spin-orbit coupling, and hence the corresponding profiles of ${\tau_{n}(k)}$ are similar. The momentum relaxation time for the 3rd and 4th electron subbands is larger than that for the 1st and 2nd  electron subbands. This is one of the reasons why we include the upper electron subbands along with the lowest ones in the calculation of the transport properties.  

To consider the process of establishment of statistical equilibrium in a system of electrons with energy ${E}$ as a result of electron collisions with lattice vibrations, we calculate the total scattering rate defined as\cite{Fischetti_PRB1988}
\begin{equation}
 \frac{1}{\tau(E)} = \frac{ \sum_{n} \int_{-\pi/2a}^{\pi/2a}  \delta \left[E - E_{n}(k)\right] / \tau_{n}(k) dk } { \sum_{n} \int_{-\pi/2a}^{\pi/2a} \delta \left[E - E_{n}(k)\right] dk}
\end{equation}

\begin{figure}[h]
\includegraphics[width=8cm]{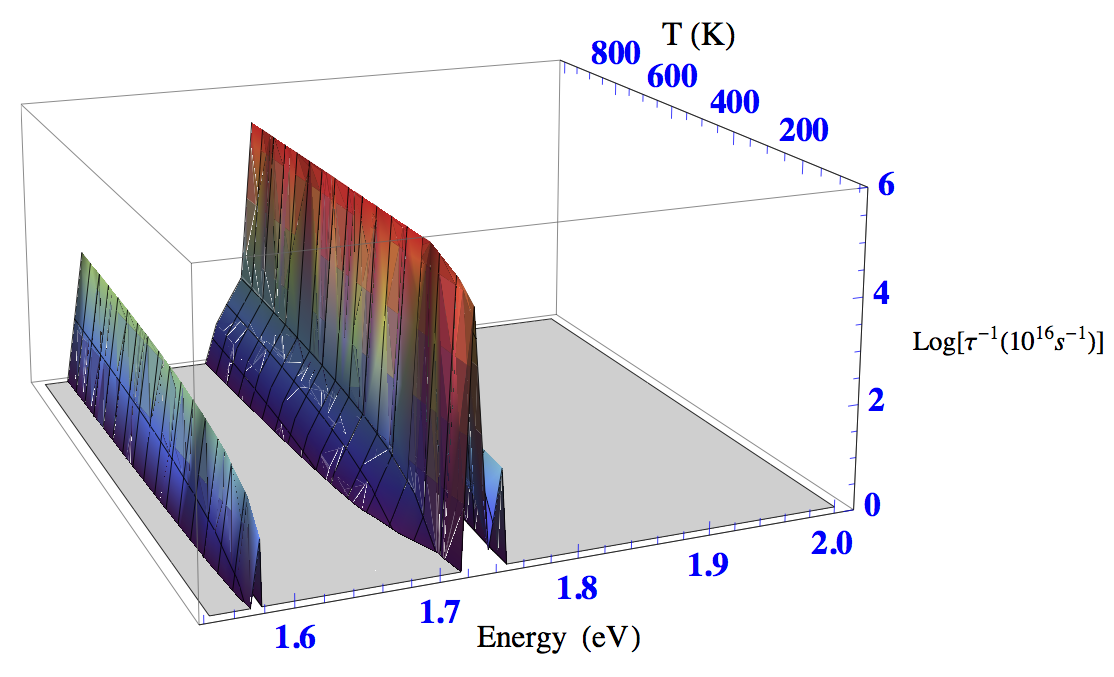}
\caption{\label{fig:fig5} (Color online) The dependence of the total scattering rate on the electron energy and temperature computed for the [100]-oriented Si NW with a thickness of 1.6 nm.}
\end{figure}

Figure~\ref{fig:fig5} represents the dependence of the total scattering rate on the electron energy and temperature for the Si NW.
The total scattering rate increases dramatically at the energies corresponding to  the bottoms of the conduction subbands. This is related to the quasi-elastic intra-subband electron scattering by acoustic phonons, which is the most important factor in the scattering. 
The population of the acoustic phonon modes increases linearly with temperature, which leads to a rather weak dependence of the scattering rate on temperature in the logarithmic plot of Fig.~\ref{fig:fig5}. When the phonon energy is comparable to the thermal energy of the electrons, the inelastic scattering processes begin to play an essential role. As a consequence, the energy domain in which the total scattering rates are large increases with temperature. 
In our calculations, both the total scattering rate and TDF do not depend on the impurity atoms concentration. 

\begin{figure}[h]
\includegraphics[width=8cm]{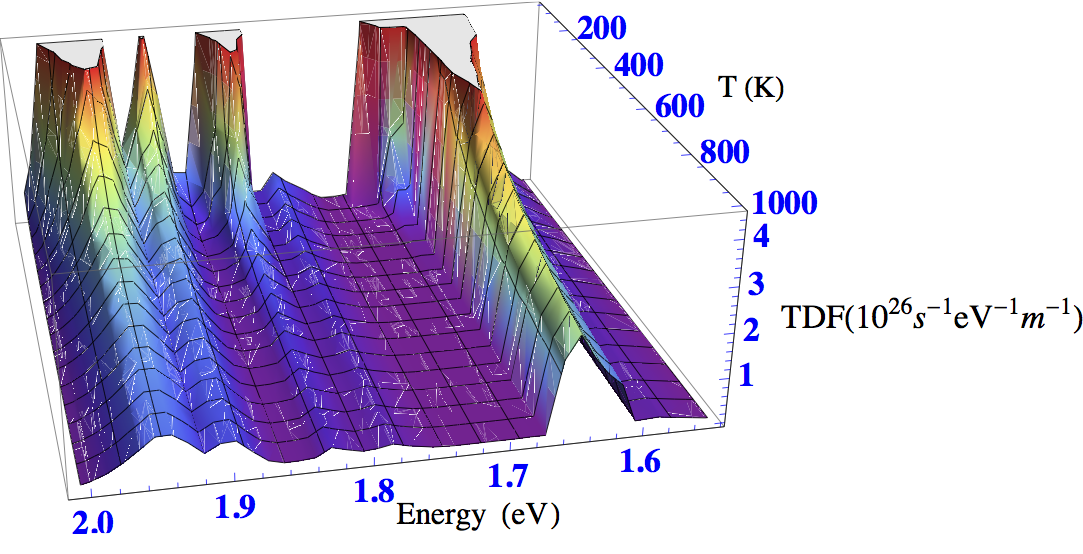}
\caption{\label{fig:fig6} (Color online) The dependence of the transport distribution function on the electron energy and temperature computed for the [100]-oriented Si NW with a thickness of 1.6 nm.}
\end{figure}

Figure~\ref{fig:fig6} depicts the dependence of the TDF on the electron energy and temperature for the Si NW. The TDF is linearly dependent on energy in the low energy region, because only few subbands contribute to the TDF. This is in agreement with the results reported by Neophytou {\it et al.}~\cite{Neophytou_PRB085313}. Both the energy and temperature dependence of the TDF is chiefly defined by the total scattering rate when the electron energy is less than 1.8 eV. In the high-energy range, the relaxation time is a rather smooth function. In this case, the electron group velocity is an essential contribution to the TDF due to the large electron band dispersion, which also leads to large relaxation times.    

\section{Temperature dependence of thermoelectric coefficients}

\begin{figure}[h]
\includegraphics[width=7cm]{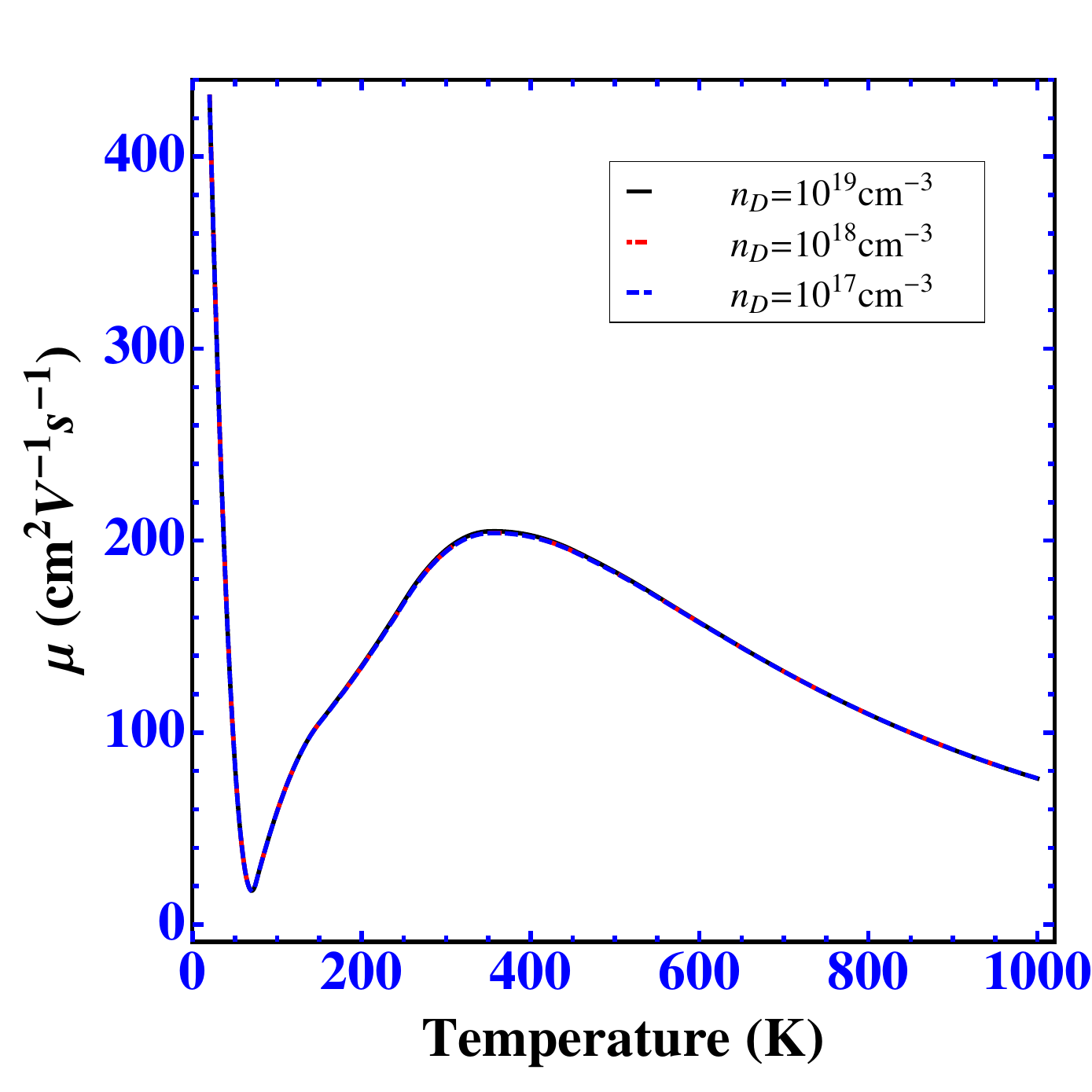}
\caption{\label{fig:fig7} (Color online) The temperature dependence of the electron mobility computed for the [100]-oriented Si NW with a thickness of 1.6 nm and dopant atoms concentration of ${10^{17}}$, ${10^{18}}$, and $10^{19}$~cm$^{-3}$.}
\end{figure}

The electronic mobility is a key factor in charge transport since it describes how the motion of an electron is affected by an applied electric field.
Figure~\ref{fig:fig7}  shows the low-field electron mobility $\mu$ as a function of temperature for the Si NW. There is a characteristic dip in the mobility in a narrow region around the temperature of 90 K. In this case, the thermal electron energy is quite small (about 8~meV) and only the lowest electron subbands contribute to the charge transport. At the same time, due to the large phonon density, the electron scattering by acoustic phonons and mixed states of acoustic and optical phonon modes increases with temperature,
resulting in a sharp drop of the mobility below 90~K. 
Further heating leads again to an increase in the electron mobility, because the number of electron subbands contributing to transport 
strongly increases whereas the efficiency of electron-phonon scattering shows only a modest increase with temperature. 
The latter is due to a reduced efficiency of scattering with acoustic phonons in the more dispersive electronic bands, as well as 
a gap in the phonon density of states between the mixed states of acoustic and optical phonons and the optical phonon modes (see Fig.~\ref{fig:fig3}).
This gap opening up at the energy of 23 meV, corresponding to a room temperature,  prevents that thermally excited optical phonons become available for scattering. 
Hence, the electron-phonon scattering remains weak in this region, and the electron mobility is relatively large (${\approx 200}$~cm$^2$V$^{-1}$s$^{-1}$). This might be beneficial for the application of thermoelectric devices based on very thin Si NWs even at room temperature.
Starting with a temperature of 400 K, the electron mobility decreases  again, as the scattering by the optical phonons increases with temperature. 
In our treatment, the mobility 
does not depend on the impurity atoms concentration. In the mid-temperature range of 300 -- 500 K, the electron mobility achieves a value of 200~cm$^2$V$^{-1}$s$^{-1}$, which is much less than the value of 1450~cm$^2$V$^{-1}$s$^{-1}$ measured for the bulk Si material.\cite{Madelung_2004} However, it is twice greater than the value obtained by Yamada {\it et al.}\cite{Yamada_JAP2012} for their Si NWs. One of the reasons may be that we considered more electron subbands in our transport calculations.
Zhang {\it et al.}\cite{Zhang_PRB2010} obtained a similar value of the electron mobility (${\approx 230}$~cm$^2$V$^{-1}$s$^{-1}$) for the $[110]$-oriented Si NW with a diameter of 1.7 nm at a room temperature.
The lower of the electron mobility in the nanowire compared to bulk is due to  (i) an increase of the electron group velocity accompanied by a decrease of effective mass,  (ii) the coincidence of electron confinement and phonon confinement, which results in an increase of the overlap of electron and phonon wave functions, and thus to an enhanced scattering rate, (iii) the lifting of the electron subband degeneracy. 
We note that a non-monotonic variation of the effective electron mobility versus temperature has also been observed experimentally in InAs NWs with a thickness of 35 nm (see Ref.~\onlinecite{Gupta_Nanotech2013}). In this study, this behavior was ascribed to Coulomb scattering from ionized surface states, but we think that other explanantions, such as the peculiar phonon spectrum of NWs, cannot be excluded. 

\begin{figure}[h]
\includegraphics[width=7.5cm]{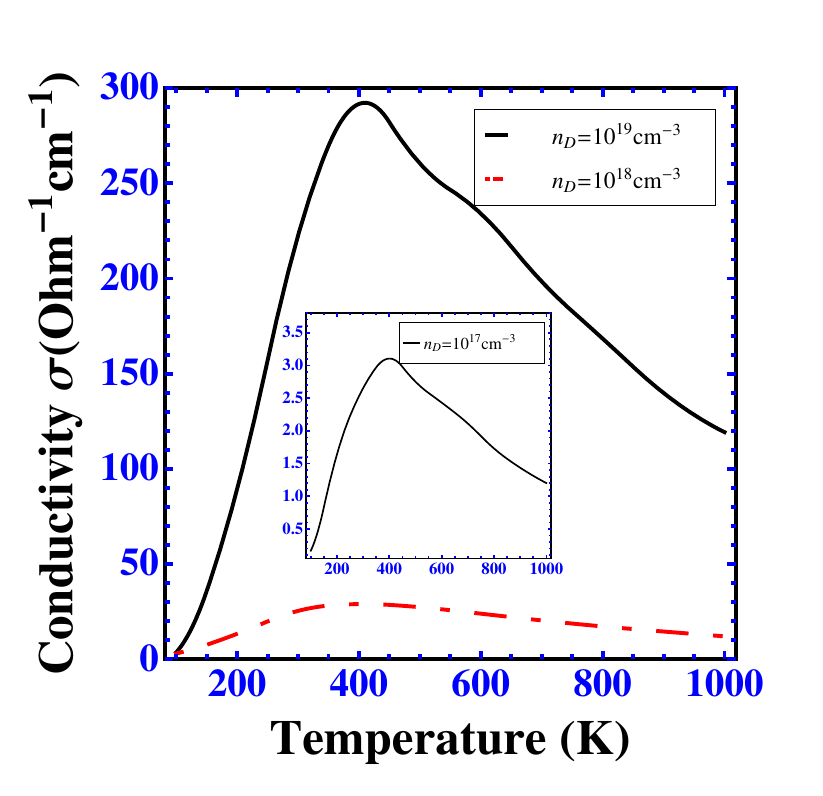}
\caption{\label{fig:fig8} (Color online) The temperature dependence of the  electrical conductivity computed for the [100]-oriented Si NW with a thickness of 1.6~nm and dopant atoms concentration of ${10^{17}}$, ${10^{18}}$ and $10^{19}$~cm$^{-3}$.}
\end{figure}

Figure~\ref{fig:fig8} displays the non-monotonic temperature dependence of the electrical conductivity $\sigma$ for the Si NW at different dopant atoms concentrations $n_{D}$. At low temperature, the electrical conductivity dramatically decreases because of the freeze-out of the electrons in the conduction subbands. It increases with temperature due to an enlargement of the electron concentration in the temperature range of the ionization regime. 
The further increase of ${\sigma}$ takes place because of enhancement of the electron mobility in the temperature range of 100--400~K. In this case, the variation of the electron concentration with temperature is much stronger than that for the electron-phonon scattering rate. That is the reason why the electrical conductivity rises monotonically at low temperature.
 At high temperature ($T>400$~K), the electron concentration weakly varies while the electron scattering by optical phonons rises; this leads to a decrease in the electrical conductivity. ${\sigma}$ achieves its maximum value of 3.10, 28.9, and 291~$\Omega^{-1}$cm$^{-1}$ at ${n_{D}}$ being equal to ${10^{17}}$, ${10^{18}}$, and $10^{19}$~cm$^{-3}$, respectively. This shows that ${\sigma}$ strongly depends on the doping concentration. The electron conductivity depends exponentially on ${T}$ in the range of low temperatures and behaves as $T^{-c}$ (with some positive constant $c$) for  $T>400$~K.
It is difficult to compare the obtained values of $\sigma$  with experimental data, because the electrical conductivity has only been measured for thicker Si NW. For example, at room temperature,  ${\sigma}$ was found to be about 125~$\Omega^{-1}$cm$^{-1}$ for a 20 nm-thick $n$-type Si NW at $n_{D}=10^{19}$~cm$^{-3}$~(see Ref.~\onlinecite{Karg_JEM2013}). In our calculations, the electron conductivity is greater than in experiment because we did not take into account the scattering of electrons by both ionized impurities and surface roughness.

\begin{figure}[h]
\includegraphics[width=7cm]{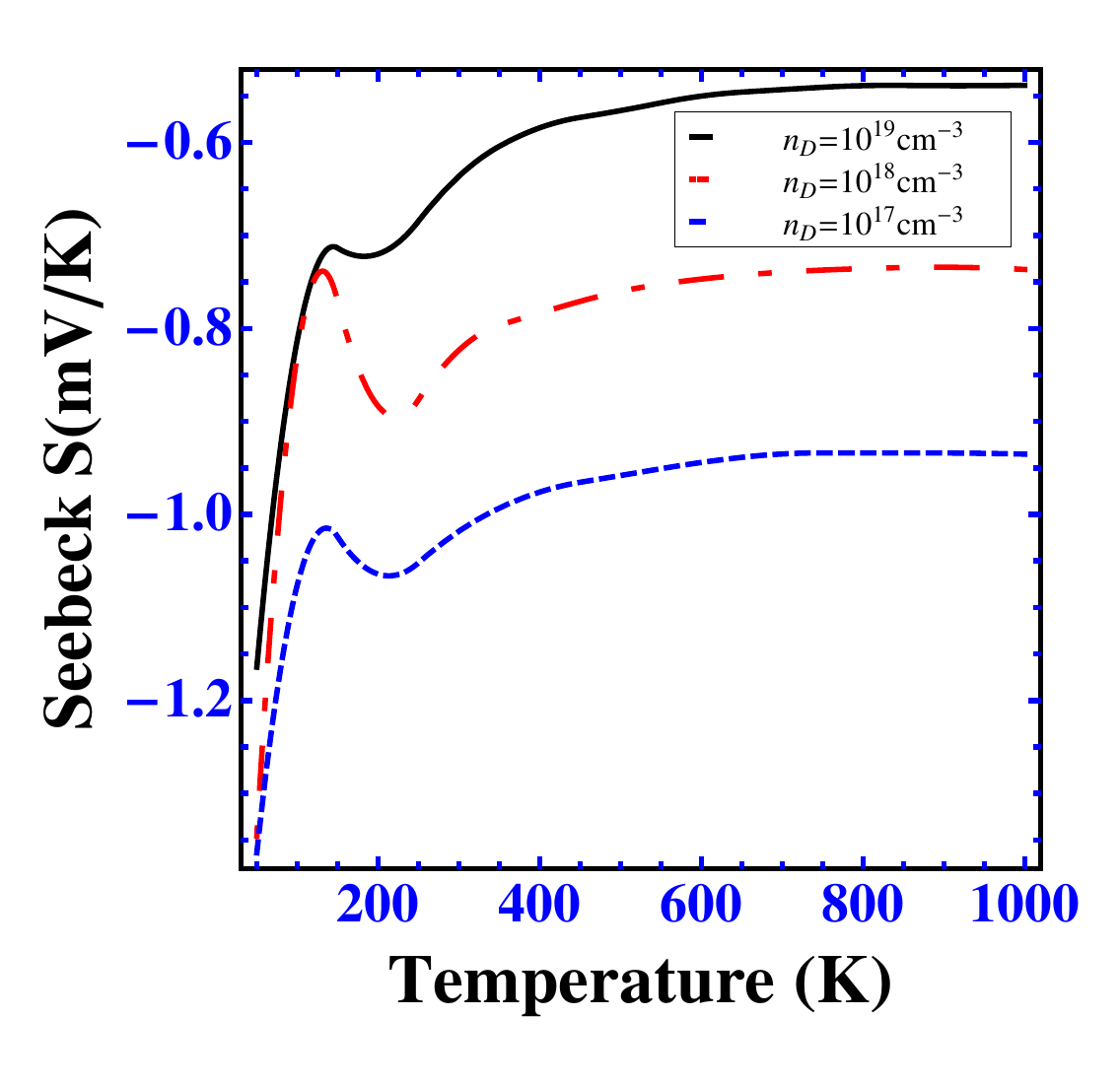}
\caption{\label{fig:fig9} (Color online) The temperature dependence of the Seebeck coefficient computed for the [100]-oriented Si NW with a thickness of 1.6~nm and dopant atoms concentration of ${10^{17}}$, ${10^{18}}$, and $10^{19}$~cm$^{-3}$.}
\end{figure}

Figure~\ref{fig:fig9} displays the non-monotonic temperature dependence of the Seebeck coefficient for the Si NW at different dopant atoms concentrations. 
As expected, the largest absolute value of $S$ is found for the lowest doping concentration. According to Eq.~(\ref{eq:Seebeck}), $S$ diverges in the limit $T\to 0$ for band-like transport, i.e. disregarding the possibility of hopping transport in an impurity band. 
The non-monotonic temperature dependence observed in $S$ and in the electron mobility have the same origin. 
As the temperature is increased above 90~K, the energy interval of the electrons that make the strongest contribution to transport shifts to higher energies. This up-shift leads to a rise in $|S|$ between 90~K and 200~K.   
The electron density of states alone cannot explain the behavior of $S$, as it appears both in the numerator and in the denominator of Eq.~(\ref{eq:Seebeck}). Hence, its effect on the Seebeck coefficient tends to cancel at least at low temperatures. 
At high temperature, the contribution of upper electron subbands becomes significant because of the weighting factor $(E-E_F)$ in the numerator [see Eq.~(\ref{eq:Seebeck})]. In this case, the Si NW is an extrinsic semiconductor and the Seebeck coefficient weakly varies with temperature. Therefore, the increase of either the electron concentration or electron-phonon scattering rate leads to the decrease of the Seebeck coefficient.

The numerical analysis shows that the variations of the electrical conductivity and of the power factor versus temperature are similar. The power factor achieves its maximum value of 0.295, 1.77, and $9.95 \times10^{-3}$ Wm$^{-1}$K$^{-2}$ at $n_{D}$ being equal to ${10^{17}}$, ${10^{18}}$, and $10^{19}$~cm$^{-3}$, correspondingly.

\begin{figure}[h]
\includegraphics[width=6cm]{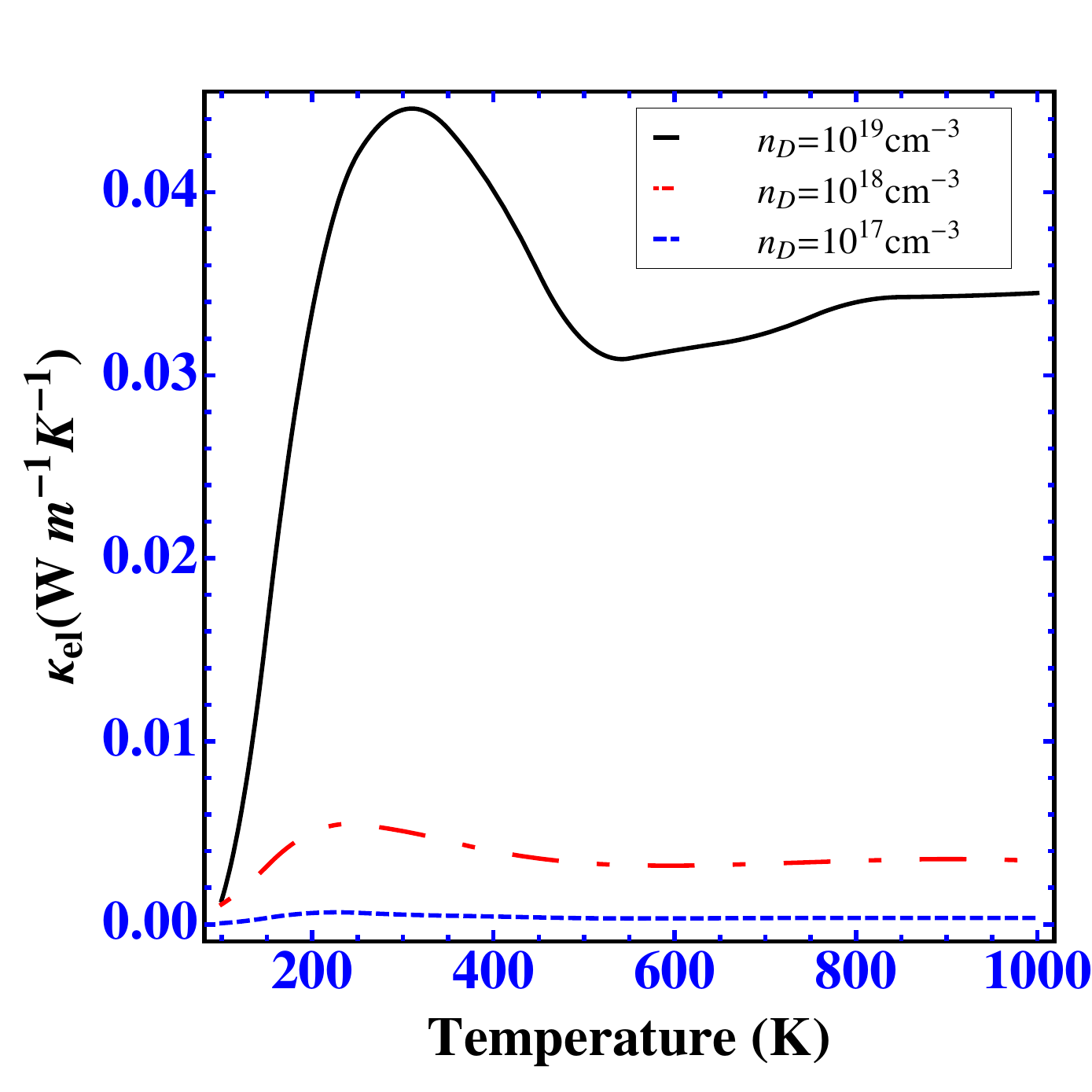}
\caption{\label{fig:fig10} (Color online) The temperature dependence of the electron thermal conductivity computed for the [100]-oriented Si NW with a thickness of 1.6 nm and dopant atoms concentration of ${10^{17}}$, ${10^{18}}$, and $10^{19}$~cm$^{-3}$.}
\end{figure}

Figure~\ref{fig:fig10} displays the non-monotonic temperature dependence of the electron thermal conductivity for the Si NW at different dopant atoms concentrations.
Both the high electron concentration and small electron-phonon scattering lead to large values of the electron thermal conductivity in the temperature range of 220--380 K.  
 While the sharp initial increase is due to the increasing number of mobile electrons, we observe that the drop in $\kappa_{el}$ at higher temperatures is not as pronounced as for $\sigma$. This indicates that excited electrons in higher subbands contribute significantly to $\kappa_{el}$. The factor $(E-E_F)^2$ appearing in $L^{(2)}$ [see Eq.~(\ref{eq:kappa_el})] as well as larger relaxation time in the higher subbands puts additional weight on the contribution of these subbands.
We can also derive this conclusion using the definition of $L^{(2)}$ given by Eq.~(\ref{eq:Lalpha_TDF}). The overlap between the transport distribution function, ${\Xi(E)}$ and the tail (corresponding to the excited electrons) of the derivative of the Fermi--Dirac distribution function, ${-\partial f_{0}(E)/\partial E}$, peaked around the Fermi level  with a width of approximately ${3.5k_{B}T}$, is significant and further increases as the Fermi level approaches the conduction subbands at higher doping concentration.   
 At the highest doping concentration of $n_D=10^{19}$~cm$^{-3}$ this even leads to an increase of $\kappa_{el}$ above 500~K. However, the electronic contribution to the thermal conductivity remains very small compared to the lattice contribution of 7~W K$^{-1}$ m$^{-1}$. 

\begin{figure}[h]
\includegraphics[width=6.5cm]{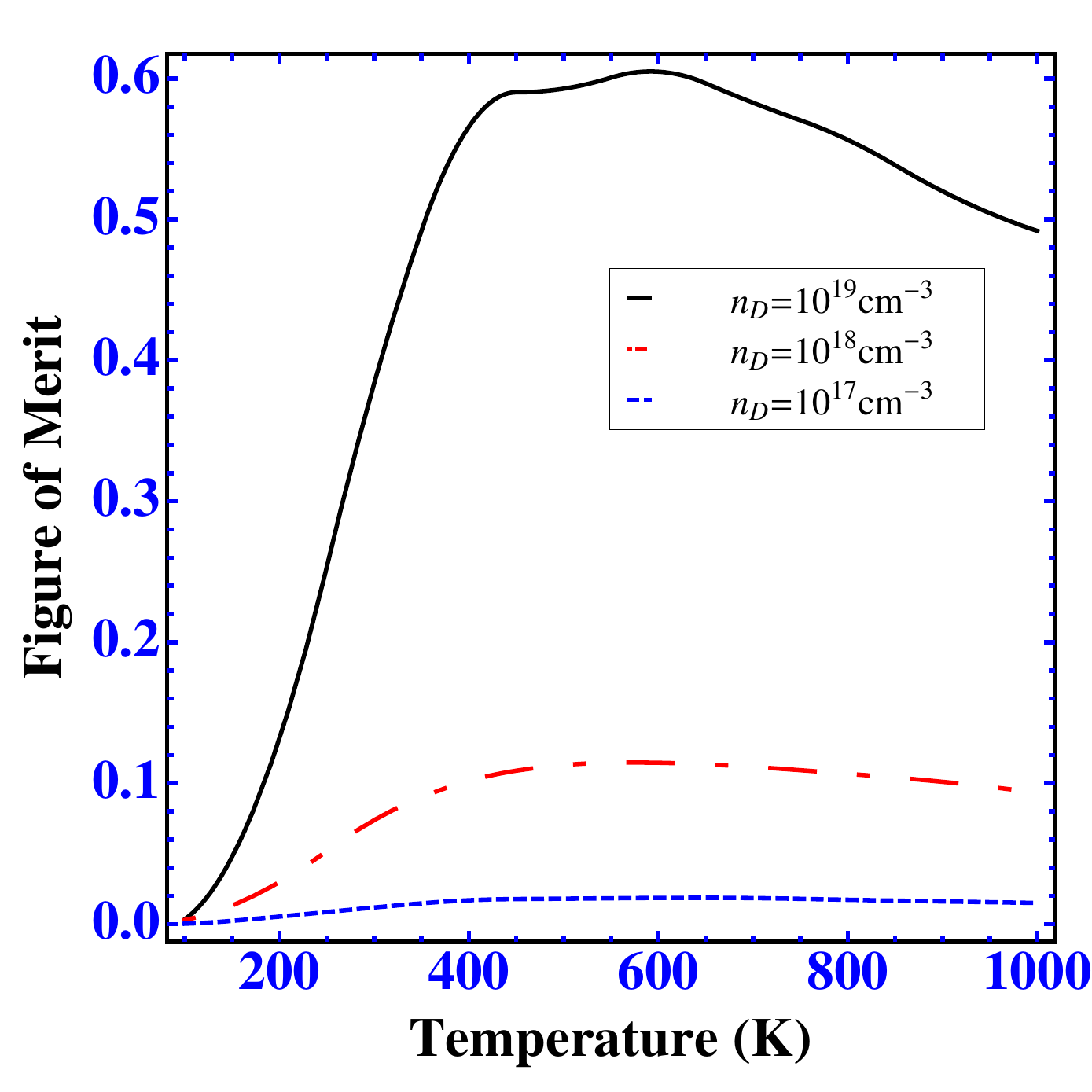}
\caption{\label{fig:fig11} (Color online) The temperature dependence of the figure of merit computed for the [100]-oriented Si NW with a thickness of 1.6 nm and dopant atoms concentration of ${10^{17}}$, ${10^{18}}$, and $10^{19}$~cm$^{-3}$.}
\end{figure}

Figure~\ref{fig:fig11} displays the temperature dependence of the figure of merit $ZT$ for the Si NW at different dopant atoms concentrations, assuming a temperature-independent $\kappa_{ph}=7$~W m$^{-1}$ K$^{-1}$. . Obviously the highest $ZT$ is reached for the highest doping concentration of $n_D=10^{19}$~cm$^{-3}$. The optimum temperature for the operation of the Si NW as thermoelectric generator is found to lie in the region of 400 -- 600~K. There, a figure of merit of 0.6 is reached. 
A comparison of the temperature dependencies of the different thermoelectric parameters lets us conclude that the temperature dependence of ${ZT}$ is mainly determined by the electrical and thermal conductivities rather than by the Seebeck coefficient. At low temperature, the variations of the figure of merit and electrical conductivity versus temperature are the same. The interval of large values of ${ZT}$ is shifted towards the higher temperatures compared to the maximum of the  electrical conductivity because of the additional factor $T$ in the definition of $ZT$.  
We note that many experimentally prepared Si NWs show surface roughness, which leads to scattering of both the electrons and phonons by the NW surface. 
While the phonon scattering reduces $\kappa_{ph}$ to a nearly temperature-independent value, and is thus advantageous for thermoelectric applications, the surface roughness scattering also lowers the mobility of the electrons. 
Moreover, we did not consider impurity scattering of the electrons. For both these reasons, the electrical conductivity to be expected in experimental nanowire samples will be lower (see, e.g., Ref.~\onlinecite{Neophytou_PRB085313}) than predicted by us. Both surface scattering and impurity scattering are independent of the lattice temperature and mostly affect electrons of low kinetic energy. Therefore we think that the temperature dependence of the thermoelectric properties predicted in our work is still meaningful even in imperfect Si NWs.

\section{Conclusions}

We have investigated the transport properties of \textit{n}-type Si NWs with a thickness of 1.6~nm 
as function of temperature by considering the atomistic electron-phonon interaction.
We extended the formulas for the first nearest-neighbor ${sp^{3}d^{5}s^{*}}$ TB model  to the more compact ${sp^3}$  second-nearest-neighbor TB model valid for semiconductor materials with the diamond crystal structure.
Our calculations show that the acoustic phonon dispersion is modified due to the influence of the nanowire surface. The surface Si-Si  dimerization results in a modification of the Si NW electronic structure. The relaxation time strongly depends on crystal momentum. The transport distribution function strongly varies with temperature and electron energy. 
The lower electron mobility in the nanowire compared to bulk Si is due to 
(i) an increase of the electron group velocity accompanied by a decrease of effective mass, 
(ii) the coincidence of electron confinement and phonon confinement, which results in an increase of the overlap of electron and phonon wave functions, and thus to an enhanced scattering rate, 
(iii) the lifting of the electron subband degeneracy. The non-monotonic temperature dependence of the mobility, of the Seebeck coefficient and of the electron thermal conductivity is owing to the highly structured 
electron and phonon 
density of states that also shows up in the transport distribution function.
This is due to the effect of the mixed states of acoustic and optical phonon modes existing in the NWs and being absent in the bulk material. The figure of merit achieves a maximum value of 0.6 at a temperature of 600 K and a donor atom concentration of  $10^{19}$~cm$^{-3}$.  
Peculiarities in the electronic and phononic structure that impede 
electron-phonon scattering are particularly important around room temperature. As a result, the electron mobility increases with temperature and becomes relatively large for such a thin nanowire (${\approx 200}$~cm$^2$V$^{-1}$s$^{-1}$). This might be beneficial for the application of Si-NW-based thermoelectric devices even at room temperature.

\begin{acknowledgments}
We thank Sung Sakong for valuable discussion of the phonon band structure  and Gregor Fiedler for providing codes to calculate the electronic band structure of Si NWs. We wish to acknowledge the Alexander von Humboldt Foundation for the funding and support of Igor Bejenari during his stay in the Duisburg-Essen University in 2012--2014. P.K. acknowledges support from the Deutsche Forschungsgemeinschaft within the Priority Programme 'Nanostructured Thermoelectrics' (SPP1386).

\end{acknowledgments}

\appendix

\section{${\textbf{sp}^3}$ 2nd nearest-neighbor TB model}

\begin{figure}[h]
\includegraphics[width=4.5cm]{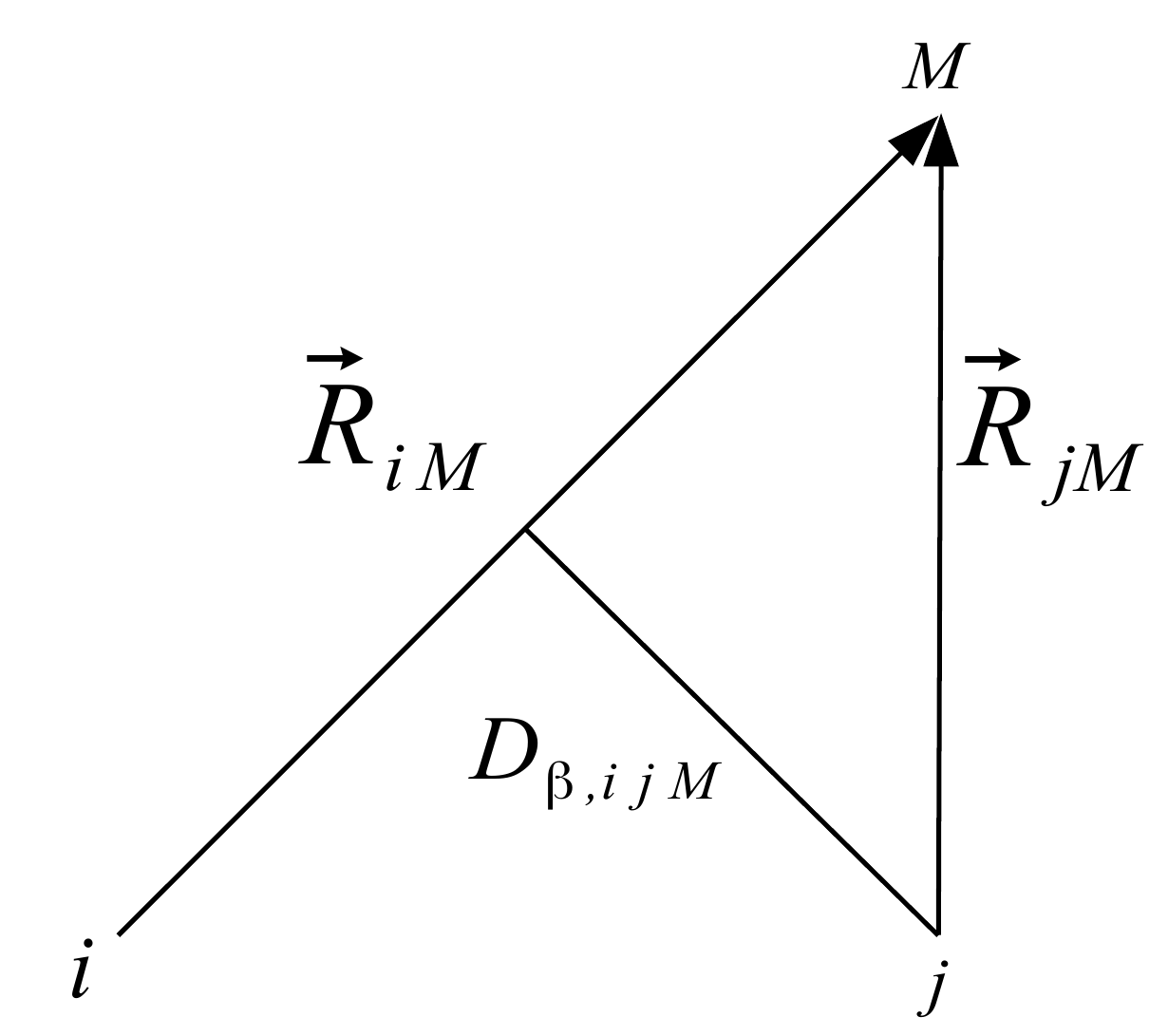}
\caption{\label{fig:fig13} Position of the first (labeled $\textit{j}$) and the second (labeled $\textit{M}$) nearest neighbors of atom $\textit{i}$.}
\end{figure}

The following formalism is valid for the diamond crystal structure. For the 1st nearest neighbors, we use the conventional Slater-Koster scheme.
The distance dependence is taken into acount via the Harrison scaling parameter  ${h_{\zeta}}$ of the ${\zeta}$th bond, and ${\vec{l_{i j}}=\vec{R_{i j}}  / R_{i j}} $ are the directional cosines. 
The Greek indices ${\beta, \gamma=x,y,z}$  are the coordinate indices of the Cartesian system. For the ease of notation, we have used the same abbreviation ${h_{\zeta}}$ for the scaling of all overlap parameters, although different values for each type of bond have been used in the actual calculations (see Ref.~\onlinecite{Grosso_PRB1995}). 
Following Refs.~\onlinecite{Slater_PR1954}, \onlinecite{santoprete:03}, and \onlinecite{Santoprete_2004}, the Hamiltonian matrix elements (transfer energy integrals) are 
\begin{equation}
{\langle s, \vec{R_{i}} | H | s, \vec{R_{j}} \rangle}^{(1)} =  V^{(1)}_{s s \sigma}(R^{0}_{i j}) \left( \frac{R^{0}_{i j}}{R_{i j}} \right)^{h_{\zeta}} 
\end{equation}
\begin{eqnarray}
\lefteqn{
{\langle s, \vec{R_i} | H | p_{\beta}, \vec{R_j} \rangle}^{(1)} =} \nonumber\\ & & = -{\langle p_{\beta}, \vec{R_i} | H | s, \vec{R_j} \rangle}^{(1)} =
  l^{\beta}_{i j} V^{(1)}_{s p \sigma} \left( \frac{R^{0}_{i j}}{R_{i j}} \right)^{h_{\zeta}} 
\end{eqnarray}
\begin{eqnarray}
\lefteqn{
{\langle p_{\beta}, \vec{R_i} \left| H \right| p_{\beta}, \vec{R_j} \rangle}^{(1)}=} \nonumber\\ & &
= \left\lbrace  \left( l^{\beta}_{i j} \right)^{2}  V^{(1)}_{p p \sigma} + \left[ 1- \left( l^{\beta}_{i j} \right)^{2} \right] V^{(1)}_{p p \pi} \right\rbrace \left( \frac{R^{0}_{i j}}{R_{i j}} \right)^{h_{\zeta}}
\end{eqnarray}
\begin{equation}
{\langle p_{\beta}, \vec{R_i} \left| H \right| p_{\gamma}, \vec{R_j} \rangle}^{(1)} =   l^{\beta}_{i j} l^{\gamma}_{i j}  \left( V^{(1)}_{p p \sigma} - V^{(1)}_{p p \pi} \right) \left( \frac{R^{0}_{i j}}{R_{i j}} \right)^{h_{\zeta}}
\end{equation}
\begin{table}[tbh]
\caption{\label{tab:table10}
The  Harrison scaling parameters and overlap parameters for Si-Si and Si-H bonds defined for the 1st and 2nd nearest neighbors. Overlap parameters ${V^{(2)}_{ss\sigma}=0}$, ${V^{(2)}_{sp\pi}=0}$ and ${V^{(2)}_{pp\sigma\pi}=0}$ for the 2nd neighbors.} 
\begin{ruledtabular}
\begin{tabular}{ccc|cc|ccc}
\textrm{Si-Si\footnote{Reference ~[\onlinecite{Grosso_PRB1995}].}} & (eV) &$h_{\zeta}$ & \textrm{Si-H\footnote{Reference ~[\onlinecite{Zheng_IEEE2005}].}}  & (eV) & \textrm{Si-Si${^{\text{a}}}$} & (eV) &$h_{\zeta}$\\
\hline
${V^{(1)}_{ss\sigma}}$& $-2.0662$ & 4.37 & ${V^{(1)}_{ss\sigma}}$ & $-3.9997$ & ${V^{(2)}_{sp\sigma}}$  & 0 & - \\
${V^{(1)}_{sp\sigma}}$& 2.085   & 3.46 & ${V^{(1)}_{sp\sigma}}$ & 4.2517 & ${V^{(2)}_{pp\sigma}}$  & 0.4444 & 7.18 \\
${V^{(1)}_{pp\sigma}}$& 3.1837   & 2.72 & ${V^{(1)}_{pp\sigma}}$ & --- & ${V^{(2)}_{pp\pi 1}}$  & 0.0844 & 8.56 \\
${V^{(1)}_{pp\pi}}$   & $-0.9488$ & 2.72 & ${V^{(1)}_{pp\pi}}$    & --- & ${V^{(2)}_{pp\pi 2}}$  & $-0.3612$ & 8.56 \\
\end{tabular}
\end{ruledtabular}
\end{table}
For the 2nd nearest neighbors, the expression for the Hamiltonian matrix elements between $s$ orbitals is analogous to the one for 1st neighbors,
\begin{equation}
{\langle s, \vec{R_{i}} | H | s, \vec{R}_{M} \rangle}^{(2)}=  V^{(2)}_{s s \sigma}\left(R^{0}_{i M}\right) \left( \frac{R^{0}_{i M}}{R_{i M}} \right)^{h_{\zeta}} \, .
\end{equation}
For the $sp$ and $pp$ interactions, however, mixing of $\sigma$ and $\pi$-type interactions, depending on the position of the 1st neighbor, needs to be taken into account. Therefore we introduce an interpolation scheme following Refs.~\onlinecite{santoprete:03} and ~\onlinecite{Santoprete_2004}. 
First we define mutually orthogonal unit vectors ${\vec{e}_{\beta}}$ pointing in the direction of the orbital ${p_{\beta}}$ at the 2nd neighbor. 
Next we introduce a unit vector ${\vec{n}_{\beta, i M}}$ that is perpendicular to the distance vector ${\vec{R}_{iM}}$ of the 2nd nearest neighbor ${M}$ with respect to atom ${i}$. This vector lies in the plane spanned by the two vectors ${\vec{e}_{\beta}}$ and ${\vec{R}_{iM}}$. 
Finally, the quantity ${D_{\beta , i j M}}$ representing the projection of the distance vector ${\vec{R}_{jM}}$ of the second nearest neighbor ${M}$ with origin at the first nearest neighbor ${j}$ of atom ${i}$ onto the  vector ${\vec{n}_{\beta, i M}}$   in used to interpolate between $\sigma$ and $\pi$-type interactions (see Fig.~\ref{fig:fig13}). 
This leads to the expressions
\begin{eqnarray}
 \vec{n}_{\beta ,i M} &=& \vec{e}_{\beta} - \frac{R_{\beta , i M} } {R^{2}_{iM} } \vec{R}_{iM} , \\
D_{\beta ,i j M} &=& \frac{ \left( \vec{R}_{j M} , \vec{n}_{\beta, i M} \right) }{\left| \vec{n}_{\beta, i M} \right|}. 
\end{eqnarray}
The 2nd neighbor matrix elements are given by 
\begin{widetext}
\begin{equation}
{\langle s, \vec{R_i} | H | p_{\beta}, \vec{R}_{M} \rangle}^{(2)}={\langle p_{\beta}, \vec{R_i} | H | s, \vec{R}_{M} \rangle}^{(2)}= \left( l^{\beta}_{i M} V^{(2)}_{s p \sigma} - \frac{4}{a} D_{\beta, i j M} V^{(2)}_{s p \pi} \right) \left( \frac{R^{0}_{i M}}{R_{i M}} \right)^{h_{\zeta}},
\end{equation}

\begin{eqnarray}
{  \langle p_{\beta}, \vec{R_i} | H | p_{\beta}, \vec{R}_{M} \rangle}^{(2)}  &= & \left[ \left( l^{\beta}_{i M} \right)^{2} V^{(2)}_{p p \sigma} + n^{2}_{\beta, i M} V^{(2)}_{p p \pi 1} \right.  \nonumber \\ 
 & + & \left. \frac{4}{a} \left| D_{\beta, i j M} \right| n_{\beta, i M} \left( V^{(2)}_{p p \pi 2} - V^{(2)}_{p p \pi 1} \right) \right]  \left( \frac{R^{0}_{i M}}{R_{i M}} \right)^{h_{\zeta}} ,
\end{eqnarray}

\begin{eqnarray}
{  \langle p_{\beta}, \vec{R_i} | H | p_{\gamma}, \vec{R}_{M} \rangle}^{(2)} &=& \left\lbrace  l^{\beta}_{i M} l^{\gamma}_{i M}  \left[ V^{(2)}_{p p \sigma} - V^{(2)}_{p p \pi 1}    
   +   \frac{2}{a} \left( \frac{D_{\beta,i j M}}{n_{\beta, i M}} +\frac{D_{\gamma,i j M}}{n_{\gamma, i M}} \right)    \left( V^{(2)}_{p p \pi 2} - V^{(2)}_{p p \pi 1} \right) \right] \right. \nonumber  \\
& + &  \left. 
 \frac{4}{a} \left( l^{\beta}_{i M} D_{\gamma, i j M} - l^{\gamma}_{i M} D_{\beta,i j M}  \right)V^{(2)}_{p p \sigma \pi}  \right\rbrace  \left(  \frac{R^{0}_{i M}}{R_{i M}} \right)^{h_{\zeta}}  ,
\end{eqnarray}
\end{widetext}
For the 1st and 2nd nearest neighbors, the  Harrison scaling parameters and overlap parameters (two-center integrals) for Si-Si and Si-H bonds are defined in Table \ref{tab:table10}. 
The diagonal matrix elements (on-site energies) of the tight binding Hamiltonian for \textit{s}- and \textit{p}-orbitals are ${E_{s}=-4.035 ~\text{eV}}$  and ${E_{p}=1.0444~\text{eV}}$  for Si-Si bonds.\cite{Grosso_PRB1995} The diagonal matrix element ${E_{s}=-1.759 ~\text{eV}}$ for the Si-H bond  has been adjusted to be compatible with the TB scheme given by Grosso {\it et al.}~\cite{Grosso_PRB1995,Zheng_IEEE2005} For the 2nd neighbors, the overlap parameters ${V^{(2)}_{pp\pi 1}}$,  ${V^{(2)}_{pp\pi 2}}$, and ${V^{(2)}_{pp\sigma\pi}}$  are defined in terms of the  transfer energy integrals as  ${V^{(2)}_{pp\pi 1} = E^{(110)}_{xx}-E^{(110)}_{xy}}$,  ${V^{(2)}_{pp\pi 2} = E^{(011)}_{xx}}$, and ${V^{(2)}_{pp\sigma \pi} = \sqrt{2}E^{(011)}_{xy}}$, correspondingly.\cite{santoprete:03} 

The matrix elements of the spin-orbit interaction between ${p}$ orbitals with a different spin are described by Kane.\cite{Kane_JPCS1956} In our calculations, we used the SO coupling parameter  ${\lambda=\Delta_{0}/3}$, where ${\Delta_{0}=0.044}$~eV is the atomic SO splitting at the ${\Gamma}$ point in the Brillouin zone.\cite{Grosso_PRB1995,Chadi_PRB1977}

\medskip

\section{Electron-phonon interaction TB Hamiltonian}

The Hamiltonian matrix elements depend on the radius vectors ${R_{i j}}$, ${R_{i M}}$, and ${R_{j M}}$. Hence, we have to consider three different kinds of spatial derivatives. For the 1st nearest neighbors, the spatial derivatives of the Hamiltonian matrix elements are given by the following expressions

\begin{widetext}
\begin{equation}
\frac{\partial{ \langle s, \vec{R_{i}} | H | s, \vec{R_{j}} \rangle}^{(1)}}{\partial \vec{R}_{ij}}= - h_{\zeta} 
\frac{V^{(1)}_{s s \sigma}(R^{0}_{i j})}{R_{i j}}  \left( \frac{R^{0}_{i j}}{R_{i j}} \right)^{h_{\zeta}} \vec{l_{ij}},
\end{equation}

\begin{equation}
\frac{\partial{ \langle s, \vec{R_{i}} | H | p_{\beta}, \vec{R_{j}} \rangle}^{(1)}}{\partial \vec{R}_{ij}}=  
\frac{V^{(1)}_{s p \sigma}}{R_{i j}}  \left( \frac{R^{0}_{i j}}{R_{i j}} \right)^{h_{\zeta}}  \left( \vec{n}_{\beta,ij} - l^{\beta}_{ij} h_{\zeta} \vec{l_{ij}} \right),
\end{equation}
\begin{eqnarray}
\frac{\partial{ \langle p_{\beta}, \vec{R_{i}} | H | p_{\beta}, \vec{R_{j}} \rangle}^{(1)}}{\partial \vec{R}_{ij}} & = &
\frac{l^{\beta}_{ij} V^{(1)}_{p p \sigma}}{R_{i j}}  \left( \frac{R^{0}_{i j}}{R_{i j}} \right)^{h_{\zeta}}  \left(2 \vec{n}_{\beta,ij} - l^{\beta}_{ij} h_{\zeta} \vec{l_{ij}} \right) \nonumber \\
& - &  \frac{ V^{(1)}_{p p \pi}}{R_{i j}}  \left( \frac{R^{0}_{i j}}{R_{i j}} \right)^{h_{\zeta}}  \left\lbrace 2 l^{\beta}_{ij} \vec{n}_{\beta,ij} + \left[1-\left(l^{\beta}_{ij}\right)^{2}\right] h_{\zeta} \vec{l_{ij}} \right\rbrace,
\end{eqnarray}

\begin{equation}
\frac{\partial{ \langle p_{\beta}, \vec{R_{i}} | H | p_{\gamma}, \vec{R_{j}} \rangle}^{(1)}}{\partial \vec{R}_{ij}}=  
\left( \frac{V^{(1)}_{p p \sigma}}{R_{i j}} -\frac{V^{(1)}_{p p \pi}}{R_{i j}} \right)  \left( \frac{R^{0}_{i j}}{R_{i j}} \right)^{h_{\zeta}}  \left( l^{\gamma}_{ij} \vec{n}_{\beta,ij} + l^{\beta}_{ij} \vec{n}_{\gamma,ij} - l^{\beta}_{ij} l^{\gamma}_{ij}  h_{\zeta} \vec{l_{ij}} \right).
\end{equation}

For the 2st neighbors, the directional derivatives of the Hamiltonian matrix elements along the radius vector ${\vec{R}_{jM}}$  are 

\begin{equation}
\frac{\partial{ \langle s, \vec{R_{i}} | H | s, \vec{R}_{M} \rangle}^{(2)}}{\partial \vec{R}_{jM}}= 0 ,
\end{equation}

\begin{equation}
\frac{\partial{ \langle s, \vec{R_{i}} | H | p_{\beta}, \vec{R}_{M} \rangle}^{(2)}}{\partial \vec{R}_{jM}}= - \frac{4}{a}  V^{(2)}_{s p \pi} \left(R^{0}_{i M} \right)  \left( \frac{R^{0}_{i M}}{R_{i M}} \right)^{h_{\zeta}} \frac{\vec{n}_{\beta,iM}}{\left| \vec{n}_{\beta,iM} \right|} ,
\end{equation}

\begin{equation}
\frac{\partial{ \langle p_{\beta}, \vec{R_{i}} | H | p_{\beta}, \vec{R}_{M} \rangle}^{(2)}}{\partial \vec{R}_{jM}}=  
\frac{4}{a} \frac{D_{\beta , i j M}}{\left| D_{\beta ,i j M} \right|} \left( V^{(2)}_{p p \pi 2} - V^{(2)}_{p p \pi 1} \right) \left( \frac{R^{0}_{i M}}{R_{i M}} \right)^{h_{\zeta}} \vec{n}_{\beta,iM},
\end{equation}

\begin{eqnarray}
\frac{\partial{ \langle p_{\beta}, \vec{R_{i}} | H | p_{\gamma}, \vec{R}_{M} \rangle}^{(2)}}{\partial \vec{R}_{jM}} & = & 
\frac{2}{a} l^{\beta}_{iM} l^{\gamma}_{iM} \left( \frac{\vec{n}_{\beta,iM}}{n^{2}_{\beta,iM}} +  \frac{\vec{n}_{\gamma,iM}}{n^{2}_{\gamma,iM}}  \right)  \left( V^{(2)}_{p p \pi 2} - V^{(2)}_{p p \pi 1} \right) \left( \frac{R^{0}_{i M}}{R_{i M}} \right)^{h_{\zeta}}   \nonumber \\
& + &  \frac{4}{a}  \left(l^{\beta}_{iM} \frac{\vec{n}_{\gamma,iM}}{\left| \vec{n}_{\gamma,iM} \right|} - l^{\gamma}_{iM}  \frac{\vec{n}_{\beta,iM}}{\left| \vec{n}_{\beta,iM} \right|}  \right)   V^{(2)}_{p p \sigma \pi}  \left( \frac{R^{0}_{i M}}{R_{i M}} \right)^{h_{\zeta}}.
\end{eqnarray}

The directional derivatives of the Hamiltonian matrix elements along the radius vector ${\vec{R}_{iM}}$  are 

\begin{equation}
\frac{\partial{ \langle s, \vec{R_{i}} | H | s, \vec{R}_{M} \rangle}^{(2)}}{\partial \vec{R}_{i M}}=  - \frac{h_{\zeta}}{R_{i M}} V^{(2)}_{s s \sigma } \left( R^{0}_{i M} \right) \left( \frac{R^{0}_{i M}}{R_{i M}} \right)^{h_{\zeta}} \vec{l}_{iM},
\end{equation}

\begin{eqnarray}
\frac{\partial{ \langle s, \vec{R_{i}} | H | p_{\beta}, \vec{R}_{M} \rangle}^{(2)}}{\partial \vec{R}_{i M}}& = & \left( V^{(2)}_{s p \sigma } \frac{\vec{n}_{\beta,iM}}{R_{i M}} - \frac{4}{a} V^{(2)}_{s p \pi } \frac{\partial D_{\beta , i j M}}{\partial \vec{R}_{i M}} \right)  \left( \frac{R^{0}_{i M}}{R_{i M}} \right)^{h_{\zeta}}
\nonumber \\
& - &   {\langle s, \vec{R_{i}} | H | p_{\beta}, \vec{R}_{M} \rangle}^{(2)} \frac{h_{\zeta}}{R_{i M}}  \vec{l}_{iM},
\end{eqnarray}

\begin{eqnarray}
\frac{\partial{ \langle p_{\beta}, \vec{R_{i}} | H | p_{\beta}, \vec{R}_{M} \rangle}^{(2)}}{\partial \vec{R}_{iM}}& = &
\frac{2 l^{\beta}_{iM}}{R_{i M}} \left( V^{(2)}_{p p \sigma} - V^{(2)}_{p p \pi 1} \right)  \left( \frac{R^{0}_{i M}}{R_{i M}} \right)^{h_{\zeta}}   \vec{n}_{\beta,iM}  \nonumber \\
& + &  \frac{4}{a} \left( V^{(2)}_{p p \pi 2} - V^{(2)}_{p p \pi 1} \right) \left( \frac{R^{0}_{i M}}{R_{i M}} \right)^{h_{\zeta}} \frac{\partial \left| D_{\beta , i j M} \right| n_{\beta,iM} }{\partial \vec{R}_{i M}} \nonumber \\
&- &   {\langle p_{\beta}, \vec{R_{i}} | H | p_{\beta}, \vec{R}_{M} \rangle}^{(2)} \frac{h_{\zeta}}{R_{i M}}  \vec{l}_{iM},
\end{eqnarray}

\begin{eqnarray}
\lefteqn{
\frac{\partial{ \langle p_{\beta}, \vec{R_{i}} | H | p_{\gamma}, \vec{R}_{M} \rangle}^{(2)}}{\partial \vec{R}_{iM}}=  
 \left[ V^{(2)}_{p p \sigma} - V^{(2)}_{p p \pi 1}    
   +   \frac{2}{a} \left( \frac{D_{\beta,i j M }}{n_{\beta, i M }} +\frac{D_{\gamma,i j M }}{n_{\gamma, i M }} \right)    \left( V^{(2)}_{p p \pi 2} - V^{(2)}_{p p \pi 1} \right) \right]  \left( \frac{R^{0}_{i M}}{R_{i M}} \right)^{h_{\zeta}}  
   } \nonumber \\
  & \times & \left( \frac{l^{\beta}_{iM}}{R_{i M}}  \vec{n}_{\gamma,iM} + \frac{l^{\gamma}_{iM}}{R_{i M}} \vec{n}_{\beta,iM} \right)   \nonumber \\
& + & \frac{2}{a} l^{\beta}_{iM } l^{\gamma}_{iM } \left( V^{(2)}_{p p \pi 2} - V^{(2)}_{p p \pi 1} \right) \left( \frac{R^{0}_{i M }}{R_{i M }} \right)^{h_{\zeta}} \left[ \frac{\partial }{\partial \vec{R}_{i M }} \left( \frac{D_{\beta , i j M }}{n_{\beta,iM }} \right) +  \frac{\partial }{\partial \vec{R}_{i M }} \left( \frac{D_{\gamma , i j M }}{n_{\gamma,iM }} \right) \right] \nonumber \\
& + & \frac{4}{a}  V^{(2)}_{p p \sigma \pi}  \left( \frac{R^{0}_{i M }}{R_{i M }} \right)^{h_{\zeta}} \left( \frac{D_{\gamma , i j M }}{R_{i M }} \vec{n}_{\beta,iM } - \frac{D_{\beta , i j M }}{R_{i M }} \vec{n}_{\gamma,iM } + l^{\beta}_{iM } \frac{\partial D_{\gamma, i j M }}{\partial \vec{R}_{i M }} - l^{\gamma}_{iM } \frac{\partial D_{\beta , i j M }}{\partial \vec{R}_{i M }}    \right) \nonumber \\
& - &  {\langle p_{\beta}, \vec{R_{i}} | H | p_{\gamma}, \vec{R}_{M} \rangle}^{(2)} \frac{h_{\zeta}}{R_{i M}}  \vec{l}_{iM}.
\end{eqnarray}

The following auxilliary quantities are used

\begin{equation}
\frac{\partial D_{\beta , i j M}}{\partial \vec{R}_{i M}} = \frac{1}{n_{\beta,iM} R_{i M}} \left[ \frac{l^{\beta}_{iM} D_{\beta , i j M} }{n_{\beta,iM}} - \frac{\left(\vec{R}_{i j}, \vec{R}_{i M} \right)}{R_{i M}} \right] \vec{n}_{\beta,iM} - \frac{l^{\beta}_{iM}}{n_{\beta,iM} R_{i M}} \vec{\Delta}_{ijM},
\end{equation}

\begin{equation}
 \frac{\partial }{\partial \vec{R}_{i M}} \left( \frac{D_{\beta , i j M}}{n_{\beta,iM}} \right) = \frac{l^{\beta}_{iM} D_{\beta , i j M}}{R_{i M} n^{3}_{\beta,iM}} \vec{n}_{\beta,iM} + \frac{1}{n_{\beta,iM }} \frac{\partial D_{\beta , i j M }}{\partial \vec{R}_{i M }},
\end{equation}

\begin{equation}
 \frac{\partial \left| D_{\beta , i j M } \right| n_{\beta,iM } }{\partial \vec{R}_{i M }} = - \frac{D_{\beta , i jM }}{\left| D_{\beta ,i jM } \right|} \frac{1}{R_{i M }} \left[ \frac{\left(\vec{R}_{i j}, \vec{R}_{i M} \right)}{R_{i M}} \vec{n}_{\beta,iM } + l^{\beta}_{iM } \vec{\Delta}_{ijM } \right],
\end{equation}

\begin{equation}
 \vec{\Delta}_{ijM } = \vec{R}_{i j} - \frac{\left(\vec{R}_{i j}, \vec{R}_{i M} \right)}{R^{2}_{i M}} \vec{R}_{i M}.
\end{equation}
\end{widetext}

The vector ${\vec{\Delta}_{ijM}}$ is perpendicular to the radius vector ${\vec{R}_{iM}}$ and lies in the plane of the two vectors ${\vec{R}_{ij}}$ and ${\vec{R}_{iM}}$. 
For silicon, using the tight-binding paramerization of Ref.~\onlinecite{Grosso_PRB1995}, the actual calculation is somewhat simpler: Not all terms in the above equations need to be evaluated because some of the overlap parameters are set equal to zero (see Table~\ref{tab:table10}).

\section{Orthomin(1) method}

To create an easy-to-read format of the iterative Orthomin(1) method, we introduce the summation-integral type operator $\hat P$ of the form
\begin{multline}
 \hat P f_{n}(k) = \frac{L_{x}}{2\pi} \sum_{n'} \int_{-\pi/2a}^{\pi/2a} S^{n,n'}(k,k') \frac{1-f_{0}(E_{n'}(k'))}{1-f_{0}(E_{n}(k))} \\
\times  v_{n'}(k') f_{n'}(k')  dk'.
\end{multline}
Using  Eq.~(\ref{eq:Dirac}), the integration in the above expression can be reduced to a summation. In terms of the summation-integral type operator and the low-temperature relaxation time~[see Eq.~(\ref{eq:tauNull})], the Boltzmann equation is written  as 
\begin{equation}
 v_{n}(k)\tau_{n}(k) - \tau_{n}^{(0)}(k) \left( \hat P\tau_{n}(k) \right) = v_{n}(k)\tau_{n}^{(0)}(k).
\label{eq:appBltzmn}
\end{equation}
 An approximate solution of Eq.~(\ref{eq:appBltzmn}) can be calculated by means of the following iterative formula 
\begin{equation}
 \tau_{n}^{(s+1)}(k)=\tau_{n}^{(s)}(k) + \alpha^{(s)} r_{n}^{(s)}(k), 
\end{equation}
 where ${r_{n}^{(s)}(k)}$ is the residual of the form
\begin{equation}
 r_{n}^{(s)}(k)=v_{n}(k) \left[ \tau_{n}^{(0)}(k) - \tau_{n}^{(s)}(k) \right] + \tau_{n}^{(0)}(k) \left(\hat P\tau_{n}^{(s)}(k) \right). 
\end{equation}
The preconditioner ${\alpha^{(s)}}$ is introduced to minimize the norm of the error. It reads
\begin{equation}
 \alpha^{(s)} = \frac{\sum_{n,k}  r_{n}^{(s)}(k) (\hat Pr_{n}^{(s)}(k))}{\sum_{n,k} \left[ \hat Pr_{n}^{(s)}(k)\right]^{2}}.
\end{equation}
The relative error defined by the formula
\begin{equation}
 \epsilon = \left\lbrace \frac{\sum_{n,k} \left[r_{n}^{(s)}(k) \right]^{2}}{\sum_{n,k} \left[  \tau_{n}^{(s)}(k) \right]^{2} } \right\rbrace^{1/2}.
\end{equation}
can be used to control the convergence of the iteration.

\providecommand{\noopsort}[1]{}\providecommand{\singleletter}[1]{#1}%

\end{document}